\documentclass[onecolumn,nofootinbib,showkeys,superscriptaddress]{revtex4}
\usepackage{amsmath,amssymb}
\usepackage{graphicx}
\usepackage{color}
\usepackage{euscript}
\usepackage{xcolor}
\usepackage{trackchanges}

\begin{document}

\title{Light propagation in a plasma on an axially symmetric and stationary spacetime: Separability of the Hamilton-Jacobi equation and shadow\footnote{Dedicated to the 60th birthday of our colleague and friend Old\v{r}ich Semer\'{a}k, who did much important work on the motion of particles in strong gravitational fields in particular around black holes.}}

\author{Barbora Bezd{\v e}kov{\' a}}
\email{barbora.bezdekova@mff.cuni.cz}
\affiliation{Institute of Theoretical Physics, Faculty of Mathematics and Physics, Charles University, Prague, Czech Republic}

\author{Volker Perlick}
\email{perlick@zarm.uni-bremen.de}
\affiliation{ZARM, University of Bremen, Bremen, Germany} 

\author{Ji\v{r}\'{i} Bi{\v c}{\' a}k}
\email{Jiri.Bicak@mff.cuni.cz}
\affiliation{Institute of Theoretical Physics, Faculty of Mathematics and Physics, Charles University, Prague, Czech Republic}

\begin{abstract}
The properties of light rays around compact objects surrounded by a plasma are affected by both strong gravitational fields described by  a general-relativistic spacetime and by a dispersive and refractive medium, characterized by the density distribution of the plasma. We study these effects employing the relativistic Hamiltonian formalism under the assumption of stationarity and axisymmetry. The necessary and sufficient conditions on the metric and   on the plasma frequency are formulated,  such that the rays   can be analytically determined from a fully separated Hamilton-Jacobi equation.   We demonstrate how these results allow to analytically calculate the photon region and the shadow, if they exist. Several specific examples are discussed in detail: the ``hairy” Kerr black holes, the Hartle-Thorne spacetime metrics, the Melvin universe, and the Teo rotating traversable wormhole. In all of these cases   a plasma medium is present as well.

\end{abstract}


\maketitle

\section{Introduction}
It was 110 years ago, as early as 1912, when Einstein, during his stay in Prague (April 1911--July 1912), sketched the basic properties of a gravitational lens in one of his notebooks, presumably on the occasion of a visit to Berlin in April to meet astronomer Erwin Freundlich.\footnote {For the details, see Ref.~\onlinecite{Renn}. Interestingly, the work of Czech astronomer F. Link was appreciated outside his country only recently although his first paper on using Einstein’s bending in photometry in Comptes Rendus was published in March 1936, more than eight months before Einstein’s paper appeared in Science Ref.~\onlinecite{EinsteinSci}. For recent discovery of Link’s work outside his country, see Ref.~\onlinecite{Link}.}

For many years gravitational lensing, one of the most lively field of research in relativistic astrophysics, has been focused on lensing by (clusters of) galaxies or microlensing by individual objects where bending angles are small. Over the last twenty years, however, more and more attention has been paid also to lensing by black holes and other compact objects where bending angles can be arbitrarily large. The most important achievement in this direction has been the observation of the ``shadow'' of the compact object -- most likely a supermassive black hole -- at the center of the galaxy M87 by the Event Horizon Telescope Collaboration \cite{EHTC19}. Related theoretical investigations are primarily aiming at distinguishing the standard Kerr black holes of general relativity from other compact objects by way of their lensing features, in particular their shadows. Among many other examples, this includes black holes in strong magnetic fields, see Refs.~\onlinecite{Melvin1,Melvin2}, or black holes with hairs and boson stars, see, e.g., Ref.~\onlinecite{Herdeiro}.  For recent reviews on shadow calculations see Refs.~\onlinecite{CunhaRev,PerlickTsupko22}.

When comparing Kerr black holes with other compact objects, in the first approach it is natural to   still assume stationarity and axisymmetry. This, however, does not give us sufficiently many symmetries for complete integrability of the equations of motion for vacuum light rays, i.e., it does not allow us to reduce the equation for lightlike geodesics to first-order form. Therefore, it is not in general possible to calculate lensing features, such as the shadow, analytically in a stationary and axisymmetric spacetime. This is possible if there is an additional constant of motion, known as a (generalized) Carter constant, which exists if and only if the Hamilton-Jacobi equation for lightlike geodesics separates. The best known example where a Carter constant exists is the Kerr metric. However, even on a spacetime for which a Carter constant exists, it is not in general true that the equations of motion for light rays \emph{in a medium} are completely integrable. In view of applications to astrophysics, the most interesting example of a medium is a plasma. A comprehensive treatment of the light propagation and formation of a shadow in   a non-magnetized and pressure-free plasma in the Kerr background is described in Ref.~\onlinecite{PerlickTsupko2017}. More recently, this approach was further studied in Refs.~\onlinecite{badia2021,konoplya2021}. In particular, the conditions on the form of the plasma frequency have been formulated which enable one to find a generalized Carter constant for, e.g., black holes
with surrounding plasma in some generalized gravity spacetimes, implying, together with stationarity and axisymmetry, the complete integrability of the equations for the rays.

The purpose of the present work is to discuss, within the framework of geometrical optics, the possibility of an \emph{analytical} approach to lensing in \emph{general} axisymmetric stationary spacetimes containing a plasma medium.   If, in addition to axisymmetry and stationarity, the existence of an equatorial plane is assumed, and if the plasma density shares all symmetries of the spacetime, complete integrability is guaranteed for light rays in the equatorial plane. However, for almost all applications one is interested also in light rays off the equatorial plane, and in some interesting spacetimes an equatorial plane does not even exist. Therefore, in this paper we want to derive the necessary and sufficient conditions for the existence of a Carter constant for light rays in a plasma on an arbitrary stationary and axisymmetric spacetime, without any further restrictions.

There are powerful, systematic studies of the separability of the geodesic equation and its first integrals in general pseudo-Riemanian (Lorentzian) spacetimes of any dimension $d>1$. These studies analyze ``separability structures” classified by the number of Killing vectors and Killing tensors which spacetimes possess. In general relativity Carter's discovery of separable spacetimes \cite{Carter} inspired a number of contributions on the separability of the Hamilton-Jacobi equation and the closely related problem of the separability of the wave equation. We refer, in particular, to the papers by S. Benenti and M. Francaviglia on spacetimes with two commuting Killing vector fields; for their review, see Ref.~\onlinecite{BenFra}. Recent studies of the separability of the Hamilton-Jacobi, Klein-Gordon and Dirac equations, including black holes and other related objects in higher dimensions are reviewed in Ref.~\onlinecite{Frolov}.
One of the most recent studies of the similar topic was performed in Ref.~\onlinecite{wang22}, where relativistic simulations of a massive Proca field evolving on a Kerr background were analyzed, allowing one to effectively study the effect of plasma leading to superradiance. However, we emphasize that there are two important differences between these earlier works, most notably the ones by Benenti and Francaviglia, and our's: Firstly they do not consider a plasma and secondly they require separability of the Hamilton-Jacobi equation for \emph{all} geodesics -- timelike, lightlike and spacelike.

The paper is organized as follows. In Sections II and III (and in the Appendix) we derive the conditions for the existence of the Carter constant for rays in axisymmetric stationary spacetimes containing plasma. As a preparation for calculating the shadow, in Section IV the photon regions in  axisymmetric and stationary spacetimes are analyzed by employing the Hamilton equations and requiring the first two derivatives of the photons' radial coordinate with respect to a curve parameter to vanish. The form of the shadows caused by black holes surrounded by plasma are studied in Section V.
The remaining parts of the paper present the analysis of specific examples: spacetimes of the hairy Kerr black holes are shown to enable separation in a way very similar to Kerr black holes without hair in Section VI. However, the Hartle-Thorne metrics describing slowly rotating relativistic stars with a quadrupole moment (with or without plasma) do not permit the separation (Section VII). As the third example the Melvin cylindrical universe with plasma is analyzed. Beside a general interest in this case, we wish to emphasize that it serves as an instructive example of the fact that the separation of the Hamilton-Jacobi equation depends crucially on the choice of the coordinates. Whereas in the spherical-type coordinates (used in the paper so far) the Carter constant (and thus the separability) cannot be obtained, in cylindrical-type coordinates it does work, the Carter constant and the photon region can be found (Section VIII). A detailed study is performed in Section IX: here the Teo wormhole metric with plasma is investigated, the separability of the Hamilton-Jacobi equation is shown, the condition for the existence of a spherical light ray around the wormhole is derived and the shadow caused by the Teo wormhole is found. The results for both the vacuum case and for the case when plasma is present are analyzed thoroughly, additionally, they are illustrated and characterized graphically. Short conclusions follow.

\section{Light propagation in a non-magnetized pressureless plasma}
Considering a description of the ray propagation through a refractive medium used by Synge \cite{Synge1960}, a respective Hamiltonian takes the form
\begin{equation}\label{hamiltonian}
\mathcal{H}(x^{\alpha},p_{\alpha})=\frac{1}{2}\left[g^{\beta\delta}p_{\beta}p_{\delta}-(n^2-1)(p_{\gamma}V^{\gamma})^2\right],
\end{equation}
where $p_{\alpha}$ is the canonical 4-momentum (called the frequency 4-vector in Ref.~\onlinecite{Synge1960}), $V^{\alpha}$ is the {4}-velocity of the medium, i.e.,
\begin{equation}
\omega(x^{\alpha})=-p_{\alpha}V^{\alpha},
\end{equation}
is the frequency measured in the rest system of the medium.  The rays are then the solutions to Hamilton's equations
\begin{equation}
    \dot{x}{}^{\alpha} = \dfrac{\partial \mathcal{H}}{\partial p_{\alpha}} \, , \quad
    \dot{p}{}_{\alpha} = -\dfrac{\partial \mathcal{H}}{\partial x^{\alpha}} \, , \quad
    \mathcal{H}( x, p ) = 0 \, .
    \label{eq:Ham0}
\end{equation}

For a non-magnetized pressureless plasma, the refractive index $n$ depends on the photon frequency $\omega(x^\alpha)$ in the form
\begin{equation}
n^2=1-\frac{\omega_{\mathrm{pl}}^2(x^\alpha)}{\omega^2(x^\alpha)} \, .
\end{equation}
Here $\omega _{\mathrm{pl}}$ is the plasma frequency whose square equals the electron density up to a constant factor. Then the
Hamiltonian (\ref{hamiltonian}) can be rewritten as
\begin{equation}
\mathcal{H}(x^{\alpha},p_{\alpha})=\frac{1}{2}\left[g^{\beta\delta} (x^{\alpha} ) p_{\beta}p_{\delta}+\omega_{pl}^2(x^\alpha) \right] \, ,
\label{eq:hampl}
\end{equation}
i.e., it is independent of the medium velocity.
The basic results of Synge are summarized in a physical way in Ref.~\onlinecite{BiHa}. Therein, the concept of the photons (light particles) associated with wave packets in media and the Minkowski energy-momentum tensor for electromagnetic field in a medium are also discussed. The Hamiltonian (\ref{eq:hampl}) can be derived from Maxwell's equations on a curved background with a two-fluid plasma model, see Breuer and Ehlers \cite{BreuerEhlers1980}.

The Hamiltonian (\ref{eq:hampl}) demonstrates that the light rays in  a plasma are determined by the same equations as the trajectories of massive particles with a ``spacetime-dependent mass function”. Moreover, if the plasma
density is nowhere zero we can introduce the conformally rescaled metric
\begin{equation}
    \tilde{g}{}_{\mu \nu} ( x^{\alpha}) = \omega _{\mathrm{pl}}^2 ( x^{\beta})
    g_{\mu \nu} (x ^{\gamma} ) \, , \quad
    \tilde{g}{}^{\mu \nu} ( x^{\alpha}) = \omega _{\mathrm{pl}}^{-2} ( x^{\beta})
    g^{\mu \nu} (x ^{\gamma} )
\label{eq:tildeg}
\end{equation}
and the modified Hamiltonian
\begin{equation}
\tilde{\mathcal{H}} ( x^{\alpha} , p_{\beta} ) =
\omega _{\mathrm{pl}}^{-2} ( x^{\gamma})
\mathcal{H}  (x ^{\rho} , p_{\sigma} )
=
\dfrac{1}{2} \big( \tilde{g}{}^{\mu \nu} ( x^{\tau}) p_{\mu} p_{\nu} + 1 \big)
\, .
\end{equation}
As multiplying the Hamiltonian with a nowhere vanishing function
leaves the solutions to (\ref{eq:Ham0}) invariant up to parametrization,
the solutions to (\ref{eq:Ham0}) actually coincide with the (non-affinely
parametrized) timelike geodesics of the metric (\ref{eq:tildeg}), i.e.,
the light rays in the plasma coincide with the trajectories of freely
falling massive particles for which separability of the Hamilton-Jacobi
equation was originally discussed by Carter in 1968 \cite{Carter68}.
However, as the metric (\ref{eq:tildeg}) cannot be globally introduced
if the plasma density is zero on some part of the spacetime, we will
not use it in the following.

\section{Derivation of the Carter constant in a general axially symmetric stationary spacetime in plasma}

Let us assume an axially symmetric stationary metric given in coordinates $(t , \varphi , r , \vartheta )$, with the metric coefficients independent of $t$ and $\varphi$.   It is our goal to find out under what conditions the Hamilton-Jacobi equation for light rays in a plasma separates, i.e., under what conditions a (generalized) Carter constant exists. We prove in the Appendix that the following two statements are true: Separation is possible only if the plasma density is independent of $t$ and $\varphi$, and if the Hamilton-Jacobi equation separates at all, then it separates in coordinates in which the metric coefficients $g_{tr}$, $g_{t \vartheta}$, $g_{\varphi r}$, $g_{\varphi \vartheta}$ and $g_{r \vartheta}$ vanish, so we may write the metric in the following form:
\begin{equation}\label{metrika}
ds^2 =-A(r,\vartheta)dt^2  + B(r,\vartheta)dr^2+2P(r,\vartheta) dtd\varphi+D(r,\vartheta)d\vartheta^2+C(r,\vartheta) d\varphi^2 \, .
\end{equation}
We assume that the Killing vector fields $\partial / \partial t$ and $\partial / \partial \varphi$ span timelike surfaces, which requires $AC+P^2>0$, $B>0$ and $D>0$. The metric coefficients may depend on arbitrarily many parameters; for example, for a Kerr black hole they depend on a mass parameter and on a spin parameter. We emphasize that the question of whether or not the Hamilton-Jacobi equation for light rays separates is a purely local question, i.e., the range of the coordinates is quite irrelevant. In particular, it is irrelevant that $t$ runs over all of $\mathbb{R}$ and $\varphi$ runs over a circle. In this sense, the following consideration is not restricted to spherical coordinates; what matters is that we have two commuting Killing vector fields that span timelike surfaces. In the following we stick with the $(t,  \varphi, r , \vartheta )$ notation, but we emphasize that the coordinates could have any meaning. It turns out that standard spherical coordinates can be inappropriate for finding the Carter constant, while in another framework respecting symmetries of the given spacetime it is quite straightforward. This aspect is further discussed in detail in Example 3.

Non-vanishing terms of the inverse metric to (\ref{metrika}) are
\begin{align}
  g^{rr} &= \frac{1}{B(r,\vartheta)},\quad g^{\vartheta \vartheta} = \frac{1}{D(r,\vartheta)}, \quad g^{\varphi \varphi} = \frac{A(r,\vartheta)}{A(r,\vartheta)C(r,\vartheta)+P^2(r,\vartheta)}, \\
  g^{tt} &= \frac{-C(r,\vartheta)}{A(r,\vartheta)C(r,\vartheta)+P^2(r,\vartheta)}, \quad g^{t\varphi} = \frac{P(r,\vartheta)}{A(r,\vartheta)C(r,\vartheta) + P^2(r,\vartheta)}.\nonumber
\end{align}

As proven in the Appendix, separation can hold only if the plasma frequency $\omega_{pl}^2(x^\alpha)$ is a function solely of the coordinates $r$ and $\vartheta$, i.e., $\omega_{pl}^2(r,\vartheta)$. In a spacetime described by the metric~(\ref{metrika}), the  Hamiltonian (\ref{eq:hampl}) then takes the form
\begin{align}
\mathcal{H}(x^\alpha,p_{\alpha})&=\frac{1}{2}\left[\frac{p_{r}^2}{B(r,\vartheta)}+\frac{p_{\vartheta}^2}{D(r,\vartheta)}+ \frac{p_{\varphi}^2A(r,\vartheta)}{A(r,\vartheta)C(r,\vartheta)+P^2(r,\vartheta)} \right.\\
&\left. -\frac{p_{t}^2C(r,\vartheta)}{A(r,\vartheta)C(r,\vartheta)+P^2(r,\vartheta)}+ \frac{2p_tp_{\varphi}P(r,\vartheta)}{A(r,\vartheta)C(r,\vartheta)+P^2(r,\vartheta)}+\omega_{pl}^2(r,\vartheta)\right].\nonumber
\end{align}

Because it obviously holds that $\frac{\partial \mathcal{H}}{\partial t}=0$ and $\frac{\partial \mathcal{H}}{\partial \varphi}=0$, we know that $p_t$ and $p_\varphi$ are constants of motion. If the spacetime is asymptotically flat, and for a light ray that reaches infinity, the component $-p_t$ is the frequency measured by a stationary observer at infinity (see, e.g., Ref.~\onlinecite{PerlickTsupko2017}). To stress its physical meaning, let us denote it as $\omega_0$. The third constant of motion in this system is $\mathcal{H}(x^\alpha,p_{\alpha})=0$. However, when applying this formula, it is useful to rewrite the Hamiltonian $\mathcal{H}(x^\alpha,p_{\alpha})$ as a function of $x^\alpha$ and $ \frac{\partial S}{\partial x^\alpha}$ to get the Hamilton-Jacobi equation. We now require the action $S$ to be separated as follows:
\begin{equation}
  S(t,\varphi,r,\vartheta)=-\omega_0t+p_\varphi\varphi+S_r(r)+S_\vartheta(\vartheta).
\end{equation}
Then, it is possible to write the Hamilton-Jacobi equation
\begin{equation}
0 = \mathcal{H} \Big( x^{\alpha} , \dfrac{\partial S}{\partial x^{\beta}} \Big)
\end{equation}
in the form
\begin{align}\label{Hamilton_Jacobi}
0&=\frac{1}{B(r,\vartheta)}\left(\frac{d S_r (r)}{d r}\right)^2+\frac{1}{D(r,\vartheta)}\left( \frac{d S_\vartheta ( \vartheta )}{d \vartheta}\right)^2+ \frac{p_{\varphi}^2A(r,\vartheta)}{A(r,\vartheta)C(r,\vartheta)+P^2(r,\vartheta)} \\
&-\frac{\omega_0^2C(r,\vartheta)}{A(r,\vartheta)C(r,\vartheta)+P^2(r,\vartheta)}-\frac{2\omega_0p_{\varphi}P(r,\vartheta)}{A(r,\vartheta)C(r,\vartheta)+P^2(r,\vartheta)}+\omega_{pl}^2(r,\vartheta).\nonumber
\end{align}
Here it is crucial that this condition has to hold for all $p_{\varphi}$ and for all $\omega _0$. The only freedom we have is to multiply this equation with a function $F(r , \vartheta)$,
\begin{align}
0&=\frac{F(r,\vartheta)}{B(r,\vartheta)}\left( \frac{d S_r (r)}{d r}\right)^2+\frac{F(r,\vartheta)}{D(r,\vartheta)}\left( \frac{d S_\vartheta (\vartheta )}{d \vartheta}\right)^2+ F(r,\vartheta)\omega_{pl}^2(r,\vartheta)
\label{eq:sep1}\\
& + \frac{F(r,\vartheta)}{A(r,\vartheta)C(r,\vartheta)+P^2(r,\vartheta)}\left[p_{\varphi}^2A(r,\vartheta)-\omega_0^2C(r,\vartheta)-2\omega_0p_{\varphi}P(r,\vartheta)\right], \nonumber
\end{align}
where $F( r , \vartheta )$ is arbitrary except for the condition that it must not have any zeros. Separability holds if and only if the right-hand side of (\ref{eq:sep1}) is, for all $p_{\varphi}$ and all $\omega _0$, the sum of a function of $r$ alone and a function of $\vartheta$ alone. We first consider the terms that are independent of $p_{\varphi}$ and $\omega _0$. As for generic light rays $dS_r(r)/dr$ and $dS_{\vartheta} ( \vartheta )/d \vartheta$ are non-zero, we find that separability can hold only if
\begin{gather}\label{podm_BD}
  \frac{F(r,\vartheta)}{B(r,\vartheta)}\equiv \mathcal{F}(r)\quad  \mathrm{and} \quad  \frac{F(r,\vartheta)}{D(r,\vartheta)}\equiv \mathcal{G}(\vartheta),
\end{gather}
which implies
\begin{gather}
   \frac{B(r,\vartheta)}{D(r,\vartheta)}=\frac{\mathcal{G}(\vartheta)}{\mathcal{F}(r)}.
   \label{eq:B/D}
\end{gather}

This is the first important condition for separability we have found: If the quotient of $B(r , \vartheta )$ and $D(r , \vartheta )$ is not of this form, we know that separability cannot hold, neither for light rays in vacuum nor in any plasma density. If (\ref{eq:B/D}) does hold, we get our function $F(r , \vartheta )$ from the first or equivalently from the second equation in (\ref{podm_BD}). As $\mathcal{G}(\vartheta)$ and $\mathcal{F}(r)$ are unique up to a common non-zero constant factor, $F(r , \vartheta )$ is fixed up to a non-zero constant factor. Here it is crucial to note that $F( r , \vartheta )$ is determined by the metric coefficients, i.e., that it is independent of the plasma. As we assume that $B(r , \vartheta)$ and $D(r , \vartheta )$ are positive, it is also clear that $\mathcal{F} (r)$ and $\mathcal{G} ( \vartheta)$ can be chosen both positive, hence $F(r , \vartheta )$ is positive and it is guaranteed that it has, indeed, no zeros.

Plugging conditions (\ref{podm_BD}) into (\ref{Hamilton_Jacobi}) yields
\begin{align}
0&=\mathcal{F}(r)\left( \frac{d S_r(r)}{d r}\right)^2+ \mathcal{G}(\vartheta)\left( \frac{d S_\vartheta ( \vartheta )}{d \vartheta}\right)^2+F(r,\vartheta)\omega_{pl}^2(r,\vartheta)
\label{eq:sep2}
\\
&+\frac{F(r,\vartheta)}{A(r,\vartheta)C(r,\vartheta)+P^2(r,\vartheta)}\left(p_{\varphi}^2A(r,\vartheta)-\omega_0^2C(r,\vartheta)-2\omega_0p_{\varphi}P(r,\vartheta)\right).\nonumber
\end{align}

Looking still at the terms that are independent of $p_{\varphi}$ and $\omega _0$ we now see that the separability condition can hold only if the plasma frequency $\omega_{pl}^2(r,\vartheta)$ is of the form
\begin{equation}\label{podm_omega}
  \omega_{pl}^2(r,\vartheta)=\frac{f_r(r)+f_\vartheta(\vartheta)}{F(r,\vartheta)},
\end{equation}
where $f_r(r)$ is an arbitrary function of $r$ and $f _{\vartheta} ( \vartheta )$
is an arbitrary function of $\vartheta$. Here we have to use the positive function $F(r, \vartheta)$ that was determined in the previous step, uniquely up to a constant factor, by the metric alone.
We now consider the terms in (\ref{eq:sep2}) that are proportional to $p _{\varphi} ^2$, $\omega _0 ^2$ and $p _{\varphi} \omega _0$, respectively. We find that the separability condition requires that $A(r,\vartheta)$, $C(r,\vartheta)$, $P(r,\vartheta)$ must meet the conditions
\begin{equation}\label{podm_ACP}
  \frac{F(r,\vartheta)}{A(r,\vartheta)C(r,\vartheta)+P^2(r,\vartheta)} X(r,\vartheta)= X_r(r)+X_\vartheta(\vartheta),
\end{equation}
where $X$ stands for $A$, $C$, or $P$.

If the separability conditions are satisfied, we can rewrite (\ref{eq:sep1}) as
\begin{gather}\label{HJ_separ}
\mathcal{F}(r)\left(\frac{d S_r (r)}{d r}\right)^2+f_r(r)+p_{\varphi}^2A_r(r)-\omega_0^2C_r(r)-2\omega_0p_{\varphi}P_r(r)=\\
-\mathcal{G}(\vartheta)\left(\frac{d S_\vartheta ( \vartheta )}{d \vartheta}\right)^2-f_\vartheta(\vartheta)-p_{\varphi}^2A_\vartheta ( \vartheta )+\omega_0^2C_\vartheta(\vartheta) +2\omega_0p_{\varphi}P_\vartheta(\vartheta)\equiv -\mathcal{K}.\nonumber
\end{gather}
From this equation we read that $\mathcal{K}$ is independent of both $r$ and $\vartheta$, i.e., that it is a constant of motion. We refer to $\mathcal{K}$ as to the (generalized) Carter constant.

Two observations are crucial: If on the given spacetime the Hamilton-Jacobi equation for vacuum light rays does not separate, then it does not separate for light rays in a plasma either, whatever the plasma density may be. And if the Hamilton-Jacobi equation for vacuum light rays \emph{does} separate, then there is an entire family of plasma densities, given by (\ref{podm_omega}) with $F(r, \vartheta)$ determined by the metric but $f_r(r)$ and $f_{\vartheta} (\vartheta)$ arbitrary, such that it separates for light rays in this plasma as well.

We end this section by showing that the Carter constant is associated
with a conformal Killing tensor field, not only in vacuum but also in
a plasma. To that end it is important to realize that (\ref{HJ_separ})
gives us the Carter constant only on the hypersurface $\mathcal{H}=0$.
To extend it to the entire cotangent bundle, we define
\begin{equation}\label{eq:CarterTM}
\mathcal{K}( x^{\alpha} , p_{\beta} )
=
K^{\mu \nu} ( x ^{\alpha} ) p_{\mu} p_{\nu}
- \dfrac{1}{2} f_r(r)+ \dfrac{1}{2} f_{\vartheta} ( \vartheta )
\end{equation}
where
\[
K^{\mu \nu} ( x ^{\alpha} ) p_{\mu} p_{\nu}
=
\dfrac{1}{2} \mathcal{G} ( \vartheta ) p_{\vartheta} ^2
-
\dfrac{1}{2} \mathcal{F} (r) p_r^2
+
\dfrac{1}{2} \big( A_{\vartheta} ( \vartheta) - A_r (r) \big) p_{\varphi}^2
\]
\begin{equation}
-
\dfrac{1}{2} \big( C_{\vartheta} ( \vartheta) - C_r (r) \big) p_t^2
+
\big( P_{\vartheta} ( \vartheta) - P_r (r) \big) p_{\varphi} p_t
\, .
\end{equation}
This defines a symmetric second-rank tensor field $K^{\mu \nu} ( x^{\alpha} )$
that depends on the metric coefficients but not on the plasma density. By a
straight-forward calculation this function
$\mathcal{K}( x^{\alpha} , p_{\beta} )$ can be equivalently rewritten
in the following two ways:
\begin{equation}
\mathcal{K} ( x^{\alpha} , p_{\beta})
=
-
\mathcal{F} (r) p_r^2
-
A_r (r) p_{\varphi}^2
+
C_r (r) p_t^2
-
2 P_r (r) p_{\varphi} p_t
- f_r(r) +F(r , \vartheta ) \mathcal{H} ( x^{\alpha} , p_{\beta} )
\, ,
\end{equation}
\begin{equation}
\mathcal{K} ( x^{\alpha} , p_{\beta})
=
\mathcal{G} ( \vartheta ) p_{\vartheta} ^2
+
A_{\vartheta} ( \vartheta) p_{\varphi}^2
-
C_{\vartheta} ( \vartheta) p_t^2
+
2 P_{\vartheta} ( \vartheta) p_{\varphi} p_t
+
f_{\vartheta} ( \vartheta )
-
F(r , \vartheta ) \mathcal{H} ( x^{\alpha} , p_{\beta} )
\, .
\end{equation}
Restricting these two expressions to the hypersurface $\mathcal{H} = 0$
shows that the function $\mathcal{K}$ defined in (\ref{eq:CarterTM}) gives
us indeed the Carter constant as it was introduced in (\ref{HJ_separ}).
The fact that $\mathcal{K}$ is a constant of motion means that the
Poisson bracket $\{ \mathcal{K}, \mathcal{H} \}$ vanishes on the
hypersurface $\mathcal{H} = 0$, for every choice of $f_r (r)$
and $f_{\vartheta} ( \vartheta )$. If we choose $f _r ( r ) =0$
and $f _{\vartheta} ( \vartheta ) = 0$, we find that the Poisson
bracket $\{ K^{\mu \nu} ( x^{\alpha} ) p_{\mu} p_{\nu} ,
g^{\rho \sigma} ( x^{\beta} ) p_{\rho} p_{\sigma} \}$
vanishes on the hypersurface $g^{\rho \sigma} ( x^{\beta} ) p_{\rho} p_{\sigma}
= 0$. This demonstrates that $K^{\mu \nu} ( x ^{\alpha} )$ is a
conformal Killing tensor field of the spacetime metric.

\section{Photon region in a general axially symmetric spacetime with plasma}

For simplicity, let us further generally   write $A_r$ instead of $A_r(r)$ etc., keeping in mind that these functions depend on the argument that they carry as an index. We can now apply the relations  $\frac{d S_r}{d r}=p_r$ and $\frac{d S_\vartheta}{d \vartheta}=p_\vartheta$. Hence, one gets
\begin{gather}
\mathcal{F}(r)p_r^2=-\mathcal{K}-f_r-p_{\varphi}^2A_r+\omega_0^2C_r+2\omega_0p_{\varphi}P_r,\\
\mathcal{G}(\vartheta)p_\vartheta^2=\mathcal{K}-f_\vartheta-p_{\varphi}^2A_\vartheta+\omega_0^2C_\vartheta+2\omega_0p_{\varphi}P_\vartheta.
\end{gather}

The photon region is the set of all events through which there is a light ray that is completely contained in a hypersurface $r = \mathrm{constant}$. For the sake of brevity we will call such light rays ``spherical'' in the following, even though $r$ is not necessarily a radius coordinate. Along each spherical light ray the equations $\dot{r}=\ddot{r}=0$ have to hold, where the overdot denotes the derivative with respect to the same parameter as in Hamilton's equations. The equations of motion

\begin{equation}
  \dot{r}=\frac{\partial \mathcal{H}}{\partial p_r}=\frac{p_r}{B(r,\vartheta)} \, , \quad  \dot{\vartheta}=\frac{\partial \mathcal{H}}{\partial p_\vartheta}=\frac{p_\vartheta}{D(r,\vartheta)} \,
\end{equation}
give

\begin{equation}\label{r_dot}
  \quad B^2(r,\vartheta)\mathcal{F}(r)\dot{r}^2=-\mathcal{K}-f_r-p_{\varphi}^2A_r+\omega_0^2C_r+2\omega_0p_{\varphi}P_r
\end{equation}
and

\begin{equation}\label{vartheta_dot}
D^2(r,\vartheta)\mathcal{G}(\vartheta)\dot{\vartheta}^2=\mathcal{K}-f_\vartheta-p_{\varphi}^2A_\vartheta+\omega_0^2C_\vartheta+2\omega_0p_{\varphi}P_\vartheta.
\end{equation}
Hence, along every spherical light ray the following two equations have to hold:
\begin{align}
  0=-\mathcal{K}-f_r-p_{\varphi}^2A_r+\omega_0^2C_r+2\omega_0p_{\varphi}P_r &\equiv R(r),\\
  0=-f'_r-p_{\varphi}^2A'_r+\omega_0^2C'_r+2\omega_0p_{\varphi}P'_r &\equiv R'(r).
\end{align}
Here the derivative with respect to $r$ is denoted as $'$.
  This gives us the constants of motion $\mathcal{K}$ and $p_\varphi$ for every spherical light ray, i.e.,
\begin{gather}
p_\varphi=\frac{\omega_0P'_r}{A'_r}\left(1\pm\sqrt{1-\frac{A'_r}{\omega^2_0P^{'2}_r}\left(f'_r-\omega_0^2C'_r\right)}\right),\label{p_phi}\\
\mathcal{K}=\frac{A_r}{A'_r}\left(f'_r-\omega_0^2C'_r\right)+\omega_0^2C_r+2\frac{\omega_0^2P'_r}{A'_r}\left(P_r-\frac{A_rP'_r}{A'_r}\right)\left(1\pm\sqrt{1-\frac{A'_r}{\omega^2_0P^{'2}_r}\left(f'_r-\omega_0^2C'_r\right)}\right)-f_r.\label{K}
\end{gather}
The other set of the equations of motion takes the form
\begin{equation}\label{varphi_dot}
\dot{\varphi}=\frac{\partial \mathcal{H}}{\partial p_\varphi}=\frac{p_{\varphi}A(r,\vartheta)}{A(r,\vartheta)C(r,\vartheta)+P^2(r,\vartheta)}-\frac{\omega_0P(r,\vartheta)}{A(r,\vartheta)C(r,\vartheta)+P^2(r,\vartheta)},
\end{equation}
\begin{equation}\label{t_dot}
\dot{t}=\frac{\partial \mathcal{H}}{\partial p_t}=\frac{\omega_0C(r,\vartheta)}{A(r,\vartheta)C(r,\vartheta)+P^2(r,\vartheta)}+\frac{p_{\varphi}P(r,\vartheta)}{A(r,\vartheta)C(r,\vartheta)+P^2(r,\vartheta)}.
\end{equation}
As the left-hand side of (\ref{vartheta_dot}) cannot be negative, the
inequality
\begin{equation}
  \mathcal{K}-f_\vartheta\ge p_{\varphi}^2A_\vartheta-\omega_0^2C_\vartheta-2\omega_0p_{\varphi}P_\vartheta
\label{eq_photon_ineq}
\end{equation}
has to hold. Inserting (\ref{p_phi}) and (\ref{K}) leads to
\begin{gather}
  \frac{A_r}{A'_r}\left(\frac{f'_r}{\omega_0^2}-C'_r\right)+C_r+2\frac{P'_r}{A'_r}\left(P_r-\frac{A_rP'_r}{A'_r}\right)\left(1\pm\sqrt{1-\frac{A'_r}{P^{'2}_r}\left(\frac{f'_r}{\omega_0^2}-C'_r\right)}\right)- \frac{f_r+f_\vartheta}{\omega_0^2}\ge \nonumber\\
  -\frac{A_\vartheta}{A'_r}\left(\frac{f'_r}{\omega_0^2}-C'_r\right)-C_\vartheta+2\frac{P'_r}{A'_r}\left(\frac{A_\vartheta P'_r}{A'_r}-P_\vartheta\right)\left(1\pm\sqrt{1-\frac{A'_r}{P^{'2}_r}\left(\frac{f'_r}{\omega_0^2}-C'_r\right)}\right).\label{photon_region}
\end{gather}
At every point $(r,\vartheta)$ where this condition holds   (either for the plus or for the minus sign before the square root) there is a spherical light ray, i.e., the inequality (\ref{photon_region}) determines the photon region.

Spherical light rays may be stable or unstable with respect to perturbations in the $r$-direction. The unstable ones are particularly important because they can serve as limit curves for light rays that approach the photon region from far away. A spherical light ray is unstable if
\begin{equation}
  0<R''(r)= -f''_r-p_{\varphi}^2A''_r+\omega_0^2C''_r+2\omega_0p_{\varphi}P''_r.
\end{equation}

\section{Black hole shadow in an axially symmetric and stationary spacetime with plasma}

We will now demonstrate that the separability of the Hamilton-Jacobi equation for light rays allows us to derive an analytical formula for the boundary curve of the shadow. We will do this for the case that our spacetime describes a black hole, but we mention that the same methodology also works for some other compact objects, e.g., for wormholes, see Example 4 below.

We want to calculate the shadow for an observer located at coordinates $(r_O,\vartheta_O)$ outside of the black hole horizon. To that end we introduce an orthonormal tetrad

\begin{align}
  e_0&=\left.Y_{1}\partial_t+Y_{2}\partial_\varphi\right|_{(r_O,\vartheta_O)},\\
  e_1&=\left.\frac{1}{\sqrt{D(r,\vartheta)}}\partial_\vartheta\right|_{(r_O,\vartheta_O)} ,\\
  e_2&= \left.Y_{3}\partial_t + Y_{4}\partial_\varphi\right|_{(r_O,\vartheta_O)},\\
  e_3&=-\left.\frac{1}{\sqrt{B(r,\vartheta)}}\partial_r\right|_{(r_O,\vartheta_O)}.
\end{align}

The coefficients $Y_{1}$, $Y_{2}$, $Y_{3}$, $Y_{4}$ are chosen so that the orthonormality conditions $g(e_0,e_0)=-1$, $g(e_2,e_2)=1$, $g(e_0,e_2)=0$  hold. Their concrete form can be derived for any given metric. We assume that $e_0$ is the four-velocity of the observer. The orthonormality conditions for our general form of the metric~(\ref{metrika}) read
\begin{align}
  -A(r,\vartheta)Y_{1}^2+2P(r,\vartheta)Y_{1}Y_{2}+ C(r,\vartheta)Y_{2}^2&= -1, \label{ortho_cond1}\\
  -A(r,\vartheta)Y_{3}^2+2P(r,\vartheta)Y_{3}Y_{4}+ C(r,\vartheta)Y_{4}^2&= 1, \\
  -A(r,\vartheta)Y_{1}Y_{3}+ P(r,\vartheta)(Y_{1}Y_{4}+Y_{2}Y_{3})+C(r,\vartheta)Y_{2}Y_{4}&= 0. \label{ortho_cond2}
\end{align}
One can see that there are actually only three equations (\ref{ortho_cond1})--(\ref{ortho_cond2}) for four unknowns which means that one of the components can be chosen arbitrarily.   This reflects the fact that we can choose for the four-velocity any normalized timelike vector in the two-space spanned by $\partial _t$ and $\partial _{\varphi}$.

A tangent vector to a light ray $\lambda(s)=(r(s),\vartheta(s),\varphi(s),t(s))$ can be written as
\begin{equation}\label{lambda_parti}
  \dot{\lambda}=\dot{r}\partial_r+\dot{\vartheta}\partial_\vartheta+\dot{\varphi}\partial_\varphi+\dot{t}\partial_t \, .
\end{equation}
  Here the overdot denotes the derivative with respect to $s$ which is the parameter that is used in Hamilton's equations. At the observation event, the same tangent vector can be written as
\begin{equation}\label{lambda_obs}
  \dot{\lambda}=\left.-\alpha e_0+\beta(\sin\theta\cos\psi e_1+\sin\theta\sin\psi e_2+\cos\theta e_3)\right|_{(r_O,\vartheta_O)}.
\end{equation}
Factors $\alpha$, $\beta$ are positive. Coordinates $\theta$ and $\psi$ denote the celestial coordinates of the observer -- the colatitude and the azimuthal angle, respectively. Due to the form of the Hamiltonian (\ref{hamiltonian}) the light rays are parameterized as $g(\dot{\lambda},\dot{\lambda})=-\omega_{pl}^2$ and thus
\begin{equation}
  \alpha^2-\beta^2=\left.\omega_{pl}^2\right|_{(r_O,\vartheta_O)}.
\end{equation}

Furthermore, $\alpha$ can be derived as
\begin{gather}\label{alpha}
  \alpha=g(\dot{\lambda},e_0)=g(\dot{\lambda},Y_1\partial_t+Y_2\partial_\varphi)=Y_1(\dot{t}g_{tt}+\dot{\varphi}g_{t\varphi})+Y_2(\dot{t}g_{t\varphi}+\dot{\varphi}g_{\varphi\varphi})=Y_1(-\omega_0)+Y_2p_\varphi,
\end{gather}
and then
\begin{gather}\label{beta}
  \beta=\sqrt{\left(-Y_1\omega_0+Y_2p_\varphi\right)^2-\omega_{pl}^2}.
\end{gather}
  Here all expressions have to be evaluated at $(r_O,\vartheta_O)$. Note that our assumption $\alpha >0$ means that the light ray goes from the observer position into the past, hence $\omega _0 = - p_t$ is negative.

A general relation between celestial coordinates $\theta$, $\psi$ and constants of motion $p_\varphi$, $\mathcal{K}$ can be found, comparing factors of $\partial_r$ and $\partial_\varphi$ in (\ref{lambda_parti}) and (\ref{lambda_obs}). This yields
\begin{align}
  \dot{r} &= -\beta\cos\theta\frac{1}{\sqrt{B(r,\vartheta)}},\\
  \dot{\varphi} &=-\alpha Y_2+\beta\sin\theta\sin\psi Y_4.
\end{align}

It is now desirable to plug into these general formulae the expressions for dotted variables (\ref{r_dot}), (\ref{vartheta_dot}), (\ref{varphi_dot}), (\ref{t_dot}), and factors $\alpha$, $\beta$ (\ref{alpha}), (\ref{beta}) derived above. One then gets
\begin{gather}
  \sin\theta=\left.\left(1+\frac{\mathcal{K}+f_r+p_{\varphi}^2A_r-\omega_0^2C_r-2\omega_0p_{\varphi}P_r}{F(r,\vartheta)(\left(-Y_1\omega_0+Y_2p_\varphi\right)^2-\omega_{pl}^2)}\right)^{1/2}\right|_{(r_O,\vartheta_O)},\label{sh_th}\\
  \sin\psi=\left.\frac{(A_r+A_\vartheta+F(r,\vartheta)Y_2^2)p_{\varphi}-(P_r+P_\vartheta+F(r,\vartheta)Y_1Y_2)\omega_0}{F^{1/2}(r,\vartheta)Y_4\left[F(r,\vartheta)\left(-Y_1\omega_0+Y_2p_\varphi\right)^2+\mathcal{K}-f_\vartheta+p_{\varphi}^2A_r-\omega_0^2C_r-2\omega_0p_{\varphi}P_r\right]^{1/2}}\right|_{(r_O,\vartheta_O)}.\label{sh_psi}
\end{gather}
For discussing the shadow we have to consider all light rays that issue from the observer position into the past. If there is only one photon region outside of the horizon, and if it consists of \emph{unstable} spherical light rays, the boundary of the shadow is determined by those light rays that asymptotically approach one of these spherical light rays. As the former must have the same constants of motion $p _{\varphi}$ and $\mathcal{K}$ as the latter, we can insert (\ref{p_phi}) and (\ref{K}) into (\ref{sh_psi}) to get $\theta$ and $\psi$ as functions of the radius coordinate $r=r_p$ on which $p_{\varphi}$ and $\mathcal{K}$ depend. This gives us the boundary of the shadow on the observer's sky as a curve parametrized by $r_p$.

Minimum and maximum values of $r_p$ can be obtained from the condition $\sin\psi=\pm1$. This is achieved when
\begin{gather}
  \left.(A_r+A_\vartheta+F(r,\vartheta)Y_2^2)p_{\varphi}-(P_r+P_\vartheta+F(r,\vartheta)Y_1Y_2)\omega_0\right|_{(r_O,\vartheta_O)}=\\
  \left.\pm F(r,\vartheta)Y_4\left[F(r,\vartheta)(\left(-Y_1\omega_0+Y_2p_\varphi\right)^2-\omega_{pl}^2)+\mathcal{K}+f_r+p_{\varphi}^2A_r-\omega_0^2C_r-2\omega_0p_{\varphi}P_r\right]^{1/2}\right|_{(r_O,\vartheta_O)}.\nonumber
\end{gather}

In a plasma, the shadow depends on $\omega _0$. We have already mentioned that in the case of asymptotic flatness and for a light ray that reaches infinity $\omega _0$ is the frequency measured by a stationary observer at infinity. As we parametrize our light rays in the past-oriented direction, $\omega _0$ is negative and the positive frequency $\omega_{\mathrm{obs}}$ measured by our observer at $(r_O, \vartheta _O )$ whose four-velocity is detremined by the tetrad coefficients $Y_1$ and $Y_2$ is
\begin{equation}
  \omega_{\mathrm{obs}}=Y_1(-\omega_0)+Y_2p_\varphi.
\end{equation}
  If $Y_2=0$, all light rays with the same $\omega _0$ give the same $\omega _{\mathrm{obs}}$; this is not the case if $Y_2 \neq 0$.

\section{Example 1: The hairy Kerr metric}

To demonstrate how our general formula works, let us now apply it to the hairy Kerr metric. This is a generalized case to the Kerr metric and obtained expressions can be thus easily compared with results derived in Ref.~\onlinecite{PerlickTsupko2017}. The metric describing a generalized Kerr black hole in the Boyer-Lindquist coordinates reads (e.g., Ref.~\onlinecite{hairyKerr2})
\begin{gather}
    ds^2=-\left(1-\frac{2rM(r)}{\rho^{2}}\right)dt^2+\frac{\rho^{2}}{\Delta}dr^2+\rho^{2}d\vartheta^2-\frac{4arM(r)}{\rho^{2}}\sin^2\vartheta dtd\varphi \nonumber\\
    +\left(r^2+a^2+\frac{2a^2rM(r)}{\rho^{2}}\sin^2\vartheta\right)\sin^2\vartheta d\varphi^2,
\end{gather}
where $\Delta=r^2+a^2-2M(r)r$, $\rho^2=r^2+a^2 \cos^2\vartheta$. The Kerr metric can be obtained as a special case when $M(r)=m= \,$const.

In this case the relevant terms become
\begin{equation}
B(r,\vartheta)=\frac{\rho^{2}}{\Delta}, \quad D(r,\vartheta)=\rho^{2},
\end{equation}
\begin{equation}
F(r,\vartheta)=\rho^{2}, \quad \mathcal{F}(r)=\Delta, \quad \mathcal{G}(\vartheta)=1,
\end{equation}
\begin{equation}
  A_r=-\frac{a^2}{\Delta}, \quad A_\vartheta=\sin^{-2}\vartheta,
\end{equation}
\begin{equation}
  C_r=\frac{(r^2+a^2)^2}{\Delta}, \quad C_\vartheta=-a^{2}\sin^{2}\vartheta,
\end{equation}
\begin{equation}
  P_r=-\frac{a(r^2+a^2)}{\Delta}, \quad P_\vartheta=a.
\end{equation}
These are formally the same expressions as those obtained for the Kerr metric, but $\Delta$ differs, containing a general function $M(r)$.

Assuming that $\omega_{pl}^2(r,\vartheta)=(f_r+f_\vartheta)/F(r,\vartheta)$, and applying the formulae introduced above leads to
\begin{gather*}
\mathcal{F}(r)\left( \frac{d S_r}{d r}\right)^2+f_r+p_{\varphi}^2A_r-\omega_0^2C_r-2\omega_0p_{\varphi}P_r=\Delta\left( \frac{d S_r}{d r}\right)^2+f_r-\frac{1}{\Delta}(ap_{\varphi}+(r^2+a^2)\omega_0)^2,
\end{gather*}
and
\begin{gather*}
    -\mathcal{G}(\vartheta)\left( \frac{d S_\vartheta}{d \vartheta}\right)^2-f_\vartheta-p_{\varphi}^2A_\vartheta+\omega_0^2C_\vartheta+2\omega_0p_{\varphi}P_\vartheta=-\left( \frac{d S_\vartheta}{d \vartheta}\right)^2-f_\vartheta-\left(\frac{p_{\varphi}}{\sin\vartheta}+a\sin\vartheta\omega_0\right)^2.
\end{gather*}
The obtained expressions formally agree with relation (27) introduced in Ref.~\onlinecite{PerlickTsupko2017}.

A difference from the Kerr metric occurs in the formula for the photon region. From general expression (\ref{photon_region}) one gets
\begin{gather}
    \left[\frac{r^2\Delta}{(r-M-rM')^2}\left(1\pm\sqrt{1-\frac{f'_r(r-M-rM')}{2r^2\omega_0^2}}\right)^2-\frac{f_r+f_\vartheta}{\omega_0^2}\right]a^2\sin^2\vartheta\ge\\
    \left[\frac{1}{r-M-rM'}\left(M(a^2-r^2)+rM'(r^2+a^2)\pm r\Delta\sqrt{1-\frac{f'_r(r-M-rM')}{2r^2\omega_0^2}}\right)+a^2\sin^2\vartheta\right]^2.
\end{gather}
The Kerr case derived in Ref.~\onlinecite{PerlickTsupko2017} is indeed obtained when $M'=0$.

Let us notice that the expression for the hairy Kerr black hole shadow would formally be the same as introduced in Ref.~\onlinecite{PerlickTsupko2017} for the Kerr black hole, though $\Delta$ can be more general as introduced above. For this reason the formula for the hairy Kerr black hole shadow is not explicitly given here. Although there is a formal correspondence of the obtained expressions, the physical situation described by these two metrics can be significantly different. A natural assumption is that in a physically relevant case, the function $M(r)$ suitably decays with increasing $r$, and the matter stress tensor arises due to a non-constant $M(r)$ satisfying energy conditions.

\section{Example 2: The Hartle-Thorne metric}
In the Appendix of Ref.~\onlinecite{HartleThorne1968} a form of the Hartle-Thorne metric for the external gravitational field of a rotating star, accurate to the second order in the angular velocity, can be found, namely
\begin{align}\label{HT_metrika}
ds^2 =-\left(1-\frac{2M}{r}+\frac{2J^2}{r^4}\right)&\left\{1+2P_2(\cos\vartheta)\left[\frac{J^2}{Mr^3}\left(1+\frac{M}{r}\right)\right.\right.\\ \nonumber
&\left.\left.+\frac{5}{8}\frac{Q-J^2/M}{M^3}Q_{2}^2\left(\frac{r}{M}-1\right)\right]\right\}dt^2   \\ \nonumber +\left(1-\frac{2M}{r}+\frac{2J^2}{r^4}\right)^{-1}&\left\{1-2P_2(\cos\vartheta)\left[\frac{J^2}{Mr^3}\left(1-\frac{5M}{r}\right)\right.\right.\\ \nonumber
&\left.\left.+\frac{5}{8}\frac{Q-J^2/M}{M^3}Q_{2}^2\left(\frac{r}{M}-1\right)\right]\right\}dr^2\\ \nonumber
+r^2\left\{1+2P_2(\cos\vartheta)\left[-\frac{J^2}{Mr^3}\right.\right.&\left.\left.\left(1+\frac{2M}{r}\right)+\frac{5}{8}\frac{Q-J^2/M}{M^3}\left\langle\frac{2M}{\sqrt{r(r-2M)}}Q_{2}^1\left(\frac{r}{M}-1\right)\right.\right.\right.\\ \nonumber
&\left.\left.\left.-Q_{2}^2\left(\frac{r}{M}-1\right)\right\rangle\right]\right\}\times \left\{d\vartheta^2+\sin^{2}\vartheta \left(d\varphi-\frac{2J}{r^3}dt\right)^2\right\}.
\end{align}
Here $M$, $J$ and $Q$ are constants. $M$ determines the mass, $J$ stands for the total angular momentum and $Q$ is the quadrupole moment of the star. Function $P_2(\cos\vartheta)$ denotes the Legendre polynomial of order 2   of the argument $\mathrm{cos} \, \vartheta$, and $Q_{n}^m \left(\frac{r}{M}-1\right)$ denotes the associated Legendre functions of the second kind of the argument $\frac{r}{M}-1$.

It was observed already by Glampedakis and Babak \cite{GlampedakisBabak2006} that for the Hartle-Thorne metric in the chosen coordinates the Hamilton-Jacobi equation for geodesics separates only in the Schwarzschild case where $J=0$ and $Q=0$; in all other cases, including the Kerr case $Q = J^2/M \neq 0$, separability fails. As this is true in particular for \emph{lightlike} geodesics, it is clear that separability cannot hold in the \emph{chosen coordinates} for light rays in a plasma, whatever the plasma density may be. In this section we will rederive this result with the help of our general equations.

For a better transparency, let us introduce
\begin{align}
  A_1 &= 1-\frac{2M}{r}+\frac{2J^2}{r^4}, \quad &j = \frac{J^2}{Mr^3}, \\
  K &= \frac{5}{8}\frac{Q-J^2/M}{M^3}, \quad &j_1 = \frac{2J}{r^3},\nonumber
  \end{align}
  and write $Q_{n}^m$ instead of $Q_{n}^m \left(\frac{r}{M}-1\right)$.
Let us further denote

\begin{gather}
\EuScript{M}_A = 1+2P_2(\cos\vartheta)\left[j\left(1+\frac{M}{r}\right)+KQ_{2}^2\right],\\
\EuScript{M}_B=1-2P_2(\cos\vartheta)\left[j\left(1-\frac{5M}{r}\right)+KQ_{2}^2\right],\\
\EuScript{M}_\varphi=1+2P_2(\cos\vartheta)\left[-j\left(1+\frac{2M}{r}\right)+K\left(\frac{2M}{\sqrt{r(r-2M)}}Q_{2}^1-Q_{2}^2\right)\right].
\end{gather}
Then the metric terms introduced in a general form in (\ref{metrika}) are
\begin{align}
  A(r,\vartheta) &= A_1\EuScript{M}_A-j_1^2r^2\sin^{2}\vartheta \EuScript{M}_\varphi, \quad &B(r,\vartheta)& = A_1^{-1}\EuScript{M}_B, \\
  C(r,\vartheta) &= r^2\sin^{2}\vartheta \EuScript{M}_\varphi, \quad &D(r,\vartheta)& = r^2\EuScript{M}_\varphi, \nonumber\\
  P(r,\vartheta) &= -j_1r^2\sin^{2}\vartheta \EuScript{M}_\varphi.\nonumber
\end{align}
It can be seen that the ratio of terms $B(r,\vartheta)$ and $D(r,\vartheta)$ gives
\begin{equation}
  \frac{B(r,\vartheta)}{D(r,\vartheta)}=\frac{\EuScript{M}_B}{A_1r^2\EuScript{M}_\varphi}.
\end{equation}
  Separability requires the right-hand side to be a function of $\vartheta$ alone divided by a function of $r$ alone. From the way in which  $\EuScript{M}_B$ and $\EuScript{M}_\varphi$ depend on $r$ and $\vartheta$ we read that this is true only if $J=0$ and $Q=0$, i.e., in the Schwarzschild case. In the Kerr case $Q=J^2/M (\neq 0)$ it is possible to change to Boyer-Lindquist coordinates in which separability is well-known to hold. The explicit form of this coordinate transformation can be found in the above-mentioned paper by Glampedakis and Babak \cite{GlampedakisBabak2006}.

\section{Example 3: The Melvin universe}
In this section we discuss the specific example of the Melvin universe   which is a solution to the Einstein-Maxwell equations with a uniform magnetic field.
It was found by Bonnor \cite{Bonnor1954} and then independently rediscovered by Melvin \cite{Melvin1964}. For a detailed discussion of the geodesics in this spacetime we refer to Ref.~\onlinecite{Melvin1966}.

The metric of the Melvin universe can be written as
\begin{equation}\label{Melvin_metric}
ds^2 =\tilde{a}^2\left[(1+\rho^2)^2(-dt^2+d\rho^2+dz^2)+\rho^2(1+\rho^2)^{-2} d\varphi^2\right],
\end{equation}
where $\tilde{a}$   is a positive constant that plays the role of an overall scaling factor, $t$ is a time coordinate and $\rho$, $z$, and $\varphi$ are the usual cylindrical polar coordinates. We denote $\Lambda(\rho)\equiv1+\rho^2$.

\subsection{Separation in the spherical coordinates}
To obtain a Carter constant, the separated terms have to be found. Because our general formulae defined above are written in the spherical coordinates ($t,r,\vartheta,\varphi$), we transform the cylindrical coordinates ($t,\rho,z,\varphi$) used in~(\ref{Melvin_metric}) by putting
\begin{align}
  t=&t, \\
  \rho=&r\sin\vartheta, \\
  z=& r\cos\vartheta,  \\
  \varphi=&\varphi.
\end{align}
In the spherical coordinates the Melvin metric can be hence rewritten in the form
\begin{equation}
ds^2 =\tilde{a}^2\left[\Lambda^2(r,\vartheta)(-dt^2+dr^2+r^2d\vartheta^2)+r^2\sin^2\vartheta\Lambda^{-2}(r,\vartheta) d\varphi^2\right],
\end{equation}
where $\Lambda(r,\vartheta)=1+r^2\sin^2\vartheta$.   If we had  $\Lambda(r,\vartheta)=1$, this would be Minkowski space.

The individual metric terms, when using the notation of Section III, read
\begin{gather}
  A(r,\vartheta)=\tilde{a}^2\Lambda^2(r,\vartheta), \quad
  B(r,\vartheta)=\tilde{a}^2\Lambda^2(r,\vartheta),\\
  C(r,\vartheta)=\tilde{a}^2r^2\sin^2\vartheta\Lambda^{-2}(r,\vartheta), \quad
  D(r,\vartheta)=\tilde{a}^2r^2\Lambda^2(r,\vartheta), \quad
  P(r,\vartheta)=0.
\end{gather}
This implies
\begin{gather}
\frac{B(r,\vartheta)}{D(r,\vartheta)}=r^{-2} \quad \Rightarrow \quad
F(r,\vartheta)=\tilde{a}^2r^{2}\Lambda^2(r,\vartheta), \quad \mathcal{F}(r)=r^2, \quad \mathcal{G}(\vartheta)=1,
\end{gather}
and
\begin{gather}
A(r,\vartheta)C(r,\vartheta)+P^2(r,\vartheta)=\tilde{a}^4r^2\sin^2\vartheta,\\
 \frac{F(r,\vartheta)}{A(r,\vartheta)C(r,\vartheta)+P^2(r,\vartheta)}A(r,\vartheta)=\Lambda^4(r,\vartheta)\sin^{-2}\vartheta=\sin^6\vartheta(r^2+\sin^{-2}\vartheta)^4\ne A_r+A_{\vartheta},\\
 \frac{F(r,\vartheta)}{A(r,\vartheta)C(r,\vartheta)+P^2(r,\vartheta)}C(r,\vartheta)=r^2=C_r,\\
 \frac{F(r,\vartheta)}{A(r,\vartheta)C(r,\vartheta)+P^2(r,\vartheta)}P(r,\vartheta)=0.
\end{gather}
We see that the term $A(r,\vartheta)$ is not in a separated form, and the Carter constant cannot be obtained.

\subsection{Separation in the cylindrical coordinates}
If, however, the cylindrical-type coordinates $\rho$, $z$ in which the Melvin metric was originally given (see~(\ref{Melvin_metric})) are used instead, the separation can be performed. Following our original notation of Section III, let us introduce
\begin{gather}
  A(\rho,z)=\tilde{a}^2\Lambda^2(\rho), \quad
  B(\rho,z)=\tilde{a}^2\Lambda^2(\rho),\\
  C(\rho,z)=\tilde{a}^2\rho^2\Lambda^{-2}(\rho), \quad
  D(\rho,z)=\tilde{a}^2\Lambda^2(\rho), \quad
  P(\rho,z)=0.
\end{gather}
All these terms are solely functions of $\rho$. Proceeding like before we find
\begin{gather}
\frac{B(\rho,z)}{D(\rho,z)}=1 \quad \Rightarrow \quad
F(\rho,z)=\tilde{a}^2\Lambda^2(\rho), \quad \mathcal{F}(\rho)=1, \quad \mathcal{G}(z)=1,
\end{gather}
and
\begin{gather}
A(\rho,z)C(\rho,z)+P^2(\rho,z)=\tilde{a}^4\rho^2,\\
 A_\rho=\rho^{-2}\Lambda^4(\rho), \quad A_z=0,\\
 C_\rho=1, \quad C_z=0,\\
 P_\rho=0, \quad P_z=0.
\end{gather}

Note that these results are not unique: We can always add a constant to $X_{\rho}$ and subtract the same constant from $X_z$, where $X$ stands for $A$, $C$ or $P$.

Applying (\ref{HJ_separ}) leads to
\begin{gather}
\left( \frac{d S_\rho}{d \rho}\right)^2+f_\rho(\rho)+p_{\varphi}^2\rho^{-2}\Lambda^4(\rho)-\omega_0^2=-\left( \frac{d S_z}{d z}\right)^2-f_z(z)\equiv -\mathcal{K},
\end{gather}
which corresponds to the results introduced in Ref.~\onlinecite{Melvin1966}.

Thus, the photon region is given by the relation
\begin{gather}
\frac{\rho\Lambda(\rho)f'_\rho  (\rho )}{2(3\rho^2-1)}+\omega_0^2-f_\rho(\rho)-f_z(z)\ge0.
\end{gather}
If there is no plasma, the entire spacetime is the photon region. Note that in the Melvin universe the photon region is filled with ``cylindrical light rays'', rather than with spherical light rays. As these cylindrical light rays are not trapped within a spatially compact region, there is no meaningful notion of a ``shadow'' in the Melvin spacetime.

We mention that there are exact solutions to the Einstein-Maxwell equations that describe a Schwarzschild or a Kerr  black hole immersed in a Melvin universe, see Ernst \cite{Ernst1976} for the Schwarzschild case and Ernst and Wild \cite{ErnstWild1976} for the Kerr case. The shadow of such a black hole was recently discussed in Refs.~\onlinecite{Melvin1,Melvin2}, respectively. However, in these cases the equations for the light rays are not separable.

\section{Example 4: The Teo wormhole metric}
To show another example where our general formulae can be used, let us turn to a stationary and axisymmetric metric describing a rotating traversable wormhole obtained by Teo \cite{Teo1998}. Our general results will give us, for light rays in a plasma on a such spacetime, the necessary and sufficient conditions for separability of the Hamilton-Jacobi equation. This will allow us to analytically determine the photon region and the shadow. We note that, without a plasma and for a subclass of Teo metrics, the shadow was already calculated by Nedkova et al., see Ref.~\onlinecite{NedkovaTinchevYazadjiev2013}, cf. Ref.~\onlinecite{Shaikh2018,GyulchevEtAl2018}.

This system is described by the metric
\begin{align}\label{Teo_metrika}
ds^2 =-N^2dt^2+\left(1-\frac{b}{r}\right)^{-1}dr^2+r^2K^2d\vartheta^2+r^2K^2\sin^{2}\vartheta\left(d\varphi-\omega dt\right)^2,
\end{align}
where $N$, $K$, $b$, and $\omega$ are functions of $r$ and $\vartheta$. The given metric is supposed to be asymptotically flat. To meet this assumption, the introduced functions must at $r\rightarrow\infty$ obey
\begin{gather}
  N=1-\frac{M}{r}+\mathcal{O}\left(\frac{1}{r^2}\right), \quad K=1+\mathcal{O}\left(\frac{1}{r}\right), \\
  \frac{b}{r}=\mathcal{O}\left(\frac{1}{r}\right), \quad \omega=\frac{2J}{r^3}+\mathcal{O}\left(\frac{1}{r^4}\right).\nonumber
\end{gather}
The chosen coordinates are supposed to cover the spacetime region between the ``neck'', which is defined by the equation $b(r,\vartheta ) = r$, and infinity. On this domain, the functions $N$, $b$, and $K$ have to be strictly positive. Moreover, it is assumed that $\partial b(r, \vartheta) / \partial \vartheta \to 0$ and $b(r, \vartheta) > r \partial b(r, \vartheta) /\partial r$ if the neck is approached. Then one can glue two copies of the spacetime together at the neck to get a wormhole that connects two asymptotically flat ends.

According to our notation introduced in (\ref{metrika}), it can be easily seen that \begin{gather}
  A(r,\vartheta)=N^2-r^2K^2\omega^2\sin^{2}\vartheta, \quad  B(r,\vartheta)=\left(1-\frac{b}{r}\right)^{-1}, \\
  C(r,\vartheta)=r^2K^2\sin^{2}\vartheta, \quad  D(r,\vartheta)=r^2K^2,\quad P(r,\vartheta)=-\omega r^2K^2\sin^{2}\vartheta.\nonumber
\end{gather}
This leads to (by using (\ref{podm_BD}), (\ref{podm_ACP}))
\begin{gather}
  F(r,\vartheta)=r^2K^2, \quad \mathcal{F}(r)=r^2K^2\left(1-\frac{b}{r}\right), \quad \mathcal{G}(\vartheta)=1,\\
  A(r,\vartheta)C(r,\vartheta)+P^2(r,\vartheta)=N^2r^2K^2\sin^{2}\vartheta,\\
  A_r=-\frac{r^2K^2}{N^2}\omega^2, \quad A_\vartheta=\sin^{-2}\vartheta,\\
  C_r=\frac{r^2K^2}{N^2}, \quad C_\vartheta=0,\\
  P_r=-\frac{r^2K^2}{N^2}\omega, \quad P_\vartheta=0.
\end{gather}
For this general form we find
\begin{gather}
  A'_r=-2\frac{rK^2}{N^2}\omega^2-r^2\omega^2\left(\frac{K^2}{N^2}\right)'-\frac{r^2K^2}{N^2}(\omega^2)',\\
  C'_r=2\frac{rK^2}{N^2}+r^2\left(\frac{K^2}{N^2}\right)',\\
  P'_r=-2\frac{rK^2}{N^2}\omega-r^2\omega\left(\frac{K^2}{N^2}\right)'-\frac{r^2K^2}{N^2}\omega'.
\end{gather}

These relations show that to be able to perform the separability of variables, each of the terms
\begin{gather*}
  \frac{K}{N}, \quad K^2\left(1-\frac{b}{r}\right), \quad  \omega
\end{gather*}
must be a function of $r$ only.

When the separability condition holds, a new radial coordinate $\ell$ can be introduced, obeying
\begin{equation}
  d \ell = \pm \left(K \sqrt{1-\frac{b}{r}}\right)^{-1} dr.
\end{equation}
This coordinate describes   the radial length in a new metric obtained by a conformal transformation $g_{\mu \nu} \mapsto K^{-2}g_{\mu \nu}$. While the original Teo coordinates $(r,\theta,\phi,t)$ describe only one half of the spacetime (from an asymptote up to the neck), the new coordinates $(\ell,\theta,\phi,t)$ cover the whole spacetime, because $\ell$ runs from $-\infty$ to $\infty$ and thus from one asymptotic end to the other.   In the following, however, we will use the original Teo coordinates.

Plugging the separated terms into (\ref{HJ_separ}) gives the equation for the Carter constant in the form
\begin{gather}
r^2K^2\left(1-\frac{b}{r}\right)\left( \frac{d S_r}{d r}\right)^2+f_r(r)-p_{\varphi}^2\frac{r^2K^2}{N^2}\omega^2-\omega_0^2\frac{r^2K^2}{N^2}+2\omega_0p_{\varphi}\frac{r^2K^2}{N^2}\omega\\
=-\left( \frac{d S_\vartheta}{d \vartheta}\right)^2-f_\vartheta(\vartheta)-p_{\varphi}^2\sin^{-2}\vartheta\equiv -\mathcal{K}.\nonumber
\end{gather}

The equations for the derivatives of $S_r$ and $S_\vartheta$ can then be rewritten as follows:
\begin{gather}
N^2\left(1-\frac{b}{r}\right)\left( \frac{d S_r}{d r}\right)^2=(\omega_0-\omega p_{\varphi})^2-\frac{N^2}{r^2K^2}(\mathcal{K}+f_r(r)),\\
\left( \frac{d S_\vartheta}{d \vartheta}\right)^2=\mathcal{K}-f_\vartheta(\vartheta)-p_{\varphi}^2\sin^{-2}\vartheta.
\label{eq:Sthetaw}
\end{gather}
For the special case that the plasma density is zero and that each of the metric coefficients $N$, $b$, $K$, and $\omega$ separately depends on $r$ only, these equations have already been given by Nedkova et al. \cite{NedkovaTinchevYazadjiev2013}.

The general expressions (\ref{p_phi}), (\ref{K}) for $\mathcal{K}$ and $p_\varphi$ in this case give
\begin{align}
\omega p_\varphi&=\frac{\omega_0\left(\mathcal{Q}'+\mathcal{Q}\frac{\omega'}{\omega}\right)\pm\omega_0\sqrt{\mathcal{Q}^2\left(\frac{\omega'}{\omega}\right)^2+\frac{f_r'}{\omega_0^2}\left(\mathcal{Q}'+2\mathcal{Q}\frac{\omega'}{\omega}\right)}}{\mathcal{Q}'+2\mathcal{Q}\frac{\omega'}{\omega}},
\label{eq:wormp}\\
\mathcal{K}&=\frac{\omega_0^2\mathcal{Q}\left(\mathcal{Q}\frac{\omega'}{\omega}\mp\sqrt{\mathcal{Q}^2\left(\frac{\omega'}{\omega}\right)^2+\frac{f_r'}{\omega_0^2}\left(\mathcal{Q}'+2\mathcal{Q}\frac{\omega'}{\omega}\right)}\right)^2}{\left(\mathcal{Q}'+2\mathcal{Q}\frac{\omega'}{\omega}\right)^2}-f_r,
\label{eq:wormK}
\end{align}
where
\begin{equation*}
  \mathcal{Q}\equiv \frac{r^2K^2}{N^2}
\end{equation*}
is a function of $r$ only and $'$ denotes the derivative with respect to $r$.

Inserting these expressions for $p_\varphi$ and $\mathcal{K}$ into (\ref{eq:Sthetaw}) leads to
\begin{gather}
  \mathcal{Q}\left(\mathcal{Q}\frac{\omega'}{\omega}\mp\sqrt{\mathcal{Q}^2\left(\frac{\omega'}{\omega}\right)^2+\frac{f_r'}{\omega_0^2}\left(\mathcal{Q}'+2\mathcal{Q}\frac{\omega'}{\omega}\right)}\right)^2-\frac{f_r+f_\vartheta}{\omega_0^2}\left(\mathcal{Q}'+2\mathcal{Q}\frac{\omega'}{\omega}\right)^2 \ge \nonumber\\ \sin^{-2}\vartheta\omega^{-2}\left(\mathcal{Q}'+\mathcal{Q}\frac{\omega'}{\omega}\pm\sqrt{\mathcal{Q}^2\left(\frac{\omega'}{\omega}\right)^2+\frac{f_r'}{\omega_0^2}\left(\mathcal{Q}'+2\mathcal{Q}\frac{\omega'}{\omega}\right)}\right)^2,
\end{gather}
which is the condition for the existence of a spherical light ray around the Teo wormhole. Alternatively, this expression can be obtained by applying the general formula (\ref{photon_region}).

With $p_{\varphi}$ and $K$ expressed as functions of the radius coordinate $r=r_p$ by (\ref{eq:wormp}) and (\ref{eq:wormK}), respectively, the equations (\ref{sh_th}) and (\ref{sh_psi}) give us the boundary curve of the shadow parametrized by $r_p$:
\begin{gather}
  \sin\theta=\left.\left(\frac{\mathcal{K}-f_\vartheta}{\mathcal{Q}(\omega p_{\varphi}-\omega_0)^2-(f_r+f_\vartheta)}\right)^{1/2}\right|_{(r_O,\vartheta_O)},\\
  \sin\psi=\left.\frac{p_\varphi}{\sin\vartheta\sqrt{\mathcal{K}-f_\vartheta}}\right|_{(r_O,\vartheta_O)}.
\end{gather}

Notice that the orthonormal tetrad
\begin{align}
  e_0&=\left.\frac{1}{N} \left(\partial_t+\omega\partial_\varphi\right)\right|_{(r_O,\vartheta_O)},
  \label{eq:tw1}
  \\
  e_1&=\left.\frac{1}{rK}\partial_\vartheta\right|_{(r_O,\vartheta_O)},
  \label{eq:tw2}
  \\
  e_2&= \left.\frac{1}{rK\sin\vartheta}
  \partial_\varphi\right|_{(r_O,\vartheta_O)},
  \label{eq:tw3}
  \\
  e_3&=-\left.\left(1-\frac{b}{r}\right)^{1/2}\partial_r\right|_{(r_O,\vartheta_O)}
  \label{eq:tw4}
\end{align}
was applied in order to obtain the relations for the wormhole shadow.

As a specific example, we consider the Teo wormhole of the form
\begin{equation}
ds^2 = \Omega (r, \vartheta ) \,
\Bigg( - dt^2 + \dfrac{dr^2}{1-\dfrac{r_0^2}{r^2}} + r^2  d \vartheta ^2
+ r^2 \mathrm{sin} ^2 \vartheta \Big( d \varphi - \dfrac{2 \, a \, dt}{r^3} \Big)^2 \Bigg)
\end{equation}
with
\begin{equation}
\Omega (r, \vartheta ) =  1+ \dfrac{(4 \, a \, \mathrm{cos} \, \vartheta )^2}{r_0^3 \, r},
\end{equation}
where $r_0$ is a positive constant with the dimension of length and $a$ is a constant with the dimension of length squared.   The radius coordinate ranges from $r_0$ to infinity. The neck is situated at $r_0$ where indeed the condition $\big(r / \Omega (r, \vartheta) \big) \big( 1- r_0^2/r^2 \big) =0$ is satisfied.

Individual terms relevant for the photon region and the shadow of this object thus are
\begin{gather}
  A(r,\vartheta)= \Omega (r , \vartheta ) \left(1-\frac{4a^2}{r^4}\sin^{2}\vartheta\right), \quad  B(r,\vartheta)=  \Omega (r , \vartheta ) \left(1-\frac{r_0^2}{r^2}\right)^{-1}, \\
  C(r,\vartheta)= \Omega (r, \vartheta)  r^2\sin^{2}\vartheta, \quad  D(r,\vartheta)=  \Omega (r, \vartheta)r^2,\quad P(r,\vartheta)=-\frac{2a}{r}   \Omega (r, \vartheta) \sin^{2}\vartheta \, .
\end{gather}

\begin{figure}[!ht]
    \centering
    \begin{tabular}{ll}
(a) &  (b)\\
    \includegraphics[width=0.49\textwidth]{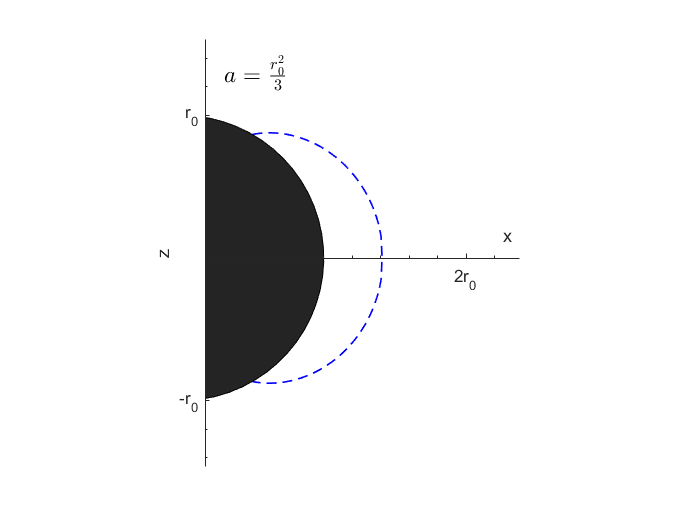} & \includegraphics[width=0.49\textwidth]{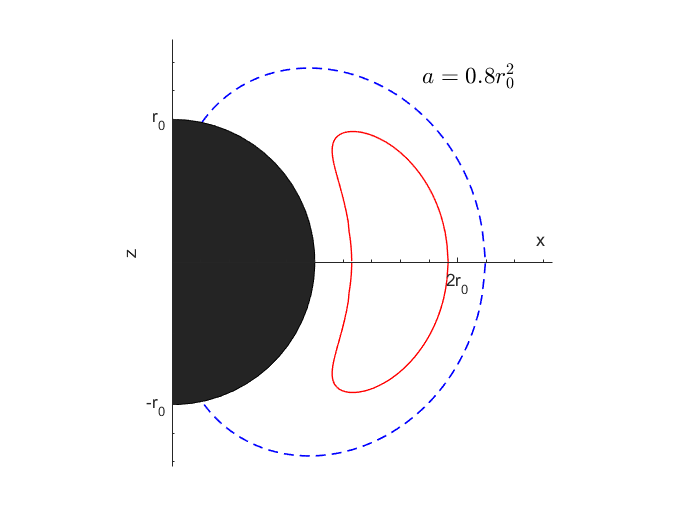}
\end{tabular}
    \caption{Photon regions for the Teo wormhole in the case when (a) $a=r_0^2/3$ and (b) $a=0.8r_0^2$. The dashed blue lines show the boundary of the photon region in the vacuum case, while the red solid line corresponds to the case that $f_r=4\omega_0^2r_0^2(r_0/r)^{1/2}$ and $f_\vartheta=0$.}
    \label{fig_01}
\end{figure}

The separated terms are
\begin{gather}
  F(r,\vartheta)= \Omega (r, \vartheta)  r^2, \quad \mathcal{F}(r)=r^2\left(1-\frac{r_0^2}{r^2}\right), \quad \mathcal{G}(\vartheta)=1,\\
  A(r,\vartheta)C(r,\vartheta)+P^2(r,\vartheta)=  \Omega ^2 (r, \vartheta ) r^2\sin^{2}\vartheta,\\
  A_r=-\frac{4a^2}{r^4}, \quad A_\vartheta=\sin^{-2}\vartheta,\\
  C_r=r^2, \quad C_\vartheta=0,\\
  P_r=-\frac{2a}{r}, \quad P_\vartheta=0.
\end{gather}
The Carter constant exists if the plasma density is of the form
\begin{equation}
\omega _{\mathrm{pl}} (r , \vartheta) ^2 =
\dfrac{f_r (r) + f_{\vartheta}(\vartheta) }{\Omega (r ,\vartheta ) r^2} \, .
\end{equation}
The equations of motion (\ref{r_dot}), (\ref{vartheta_dot}),(\ref{varphi_dot}) and (\ref{t_dot}) read:
\begin{equation}
  \Omega ^2 (r, \vartheta )  r^4\, \dot{r}{}^2 = \Big( r^2 - r_0^2 \Big)
\Bigg( - \mathcal{K} + \Big( \dfrac{2a}{r^2} \, p _{\varphi} - \omega _0 r \Big) ^2
- f_r (r) \Bigg)  \, ,
\label{eq:rdotw}
\end{equation}
\begin{equation}
  \Omega ^2 (r, \vartheta ) r^4 \, \dot{\vartheta}{}^2 =
 \mathcal{K} - \dfrac{p _{\varphi}^2}{\mathrm{sin} ^2 \vartheta}  - f_{\vartheta} \, ,
\label{eq:thdotw}
\end{equation}
\begin{equation}
\dot{\varphi} =
\dfrac{
p_{\varphi} \big( r^4-4 a^2 \mathrm{sin} ^2 \vartheta \big) + 2 \, a \, \omega _0 \, r^5 \, \mathrm{sin} ^2 \vartheta
}{
  \Omega (r, \vartheta ) \,  r^6 \, \mathrm{sin} ^2 \vartheta
} \, ,
\label{eq:phdotw}
\end{equation}
\begin{equation}
\dot{t} =
\dfrac{\omega _0 \, r^3 -2 \, a \, p_{\varphi}}{  \Omega (r, \vartheta ) \, r^3}
 \, .
\label{eq:tdotw}
\end{equation}
For spherical light rays the right-hand side of (\ref{eq:rdotw}) and its derivative must be equal to zero. Evaluating these two equations we see that they are always satisfied at the neck, at $r=r_0$, and the Carter constant of the corresponding light rays is a function of $p_{\varphi}$, i.e.,
\begin{equation}
\mathcal{K}  ( p _{\varphi} ) = \Big( \dfrac{2a}{r^2} \, p _{\varphi} - \omega _0 r \Big) ^2
- f_r (r) \, .
\label{eq:teoKp}
\end{equation}
In vacuum, these spherical light rays at the neck are unstable, but in the plasma some of them may become stable, depending on the special form of the function $f_{r} (r)$. For $r \neq r_0$, setting the right-hand side of (\ref{eq:rdotw}) and its derivative equal to zero and solving these two equations for $p_{\varphi}$ and $\mathcal{K}$ shows that for a light ray on a sphere of radius $r_p$ these constants of motion are
\begin{align}\label{eq:paramw}
    p_\varphi (r_p) &=\frac{\omega_0r_p^2}{8a}\left(r_p \mp\sqrt{9r_p^2-4r_p \frac{f'_r (r_p)}{\omega^2_0}}\right),\\
    \mathcal{K} (r_p) &=\frac{\omega^2_0}{16}\left(3r_p\pm\sqrt{9r_p^2-4r_p \frac{f'_r (r_p)}{\omega^2_0}}\right)^2-f_r (r_p) \, .
\end{align}
So in addition to the photon sphere at the neck we have in general also a photon region, given by (\ref{photon_region}) specified to the case at hand,
\begin{equation}
  \left(3r\pm\sqrt{9r^2-4r\frac{f'_r(r)}{\omega^2_0}}\right)^2- 16\frac{f_r(r)+f_\vartheta (\vartheta )}{\omega_0^2}\ge
  \frac{r^4}{4a^2 \mathrm{sin}^2 \vartheta}\left(r\mp\sqrt{9r^2-4r\frac{f'_r(r)}{\omega^2_0}}\right)^2.
\label{eq:teophreg}
\end{equation}
From (\ref{eq:rdotw}) one gets the condition for the spherical light orbits in the photon region to be unstable:
\begin{equation}
    0<R''(r)=-f_r''(r)+2\left(\frac{40a^2}{r^6}p_\varphi^2+\omega_0^2-\frac{4a}{r^3}\omega_0p_\varphi\right).
\end{equation}
In vacuum, only the upper sign in (\ref{eq:teophreg}) is possible and for $a < r_0^2/6$ the photon region does not exist, i.e., only the unstable photon orbits in the neck can serve as limit curves for light rays that determine the boundary of the shadow. For $a> r_0^2/6$ the photon region exists. It is divided by the photon sphere at the neck into two symmetric parts. The boundary curve of the shadow is partly formed by light rays that spiral towards the photon sphere and partly by light rays that spiral towards unstable spherical orbits in the photon region at radii $ r_p \neq r_0$. In a plasma, the new feature is that the photon region may become detached from the neck, with the spherical orbits at the neck being stable. The boundary curve of the shadow is then entirely determined by light rays that spiral towards spherical orbits in the component of the photon region that is on the same side of the neck as the observer.
The photon orbits for a wormhole with the choice of $a=r_0^2/3$ and $a=0.8r_0^2$, respectively, are shown in Fig.~\ref{fig_01}. Note that if the plasma profile is chosen as $f_r=4\omega_0^2r_0^2(r_0/r)^{1/2}$ and $f_\vartheta=0$, the photon region of the Teo wormhole with $a=r_0^2/3$ does not exist.

For calculating the shadow, we choose the same tetrad as in (\ref{eq:tw1})-(\ref{eq:tw4}) which now takes the following form:
\begin{align}
  e_0&=\left.\frac{1}{\sqrt{  \Omega (r, \vartheta )}}\left(\partial_t+\frac{2a}{r^3}\partial_\varphi\right)\right|_{(r_O,\vartheta_O)},\\
  e_1&=\left.\frac{1}{\sqrt{  \Omega (r, \vartheta )}r}\partial_\vartheta\right|_{(r_O,\vartheta_O)} ,\\
  e_2&= \left. \frac{1}{\sqrt{  \Omega (r, \vartheta )}r\sin\vartheta}\partial_\varphi\right|_{(r_O,\vartheta_O)},\\
  e_3&=-\left.\frac{1}{\sqrt{  \Omega (r, \vartheta )}}\left(1-\frac{r_0^2}{r^2}\right)^{1/2}\partial_r\right|_{(r_O,\vartheta_O)} \, .
\end{align}
Here the observer position $r_O$ should not be confused with the radius $r_0$ of the neck.
Comparing the coefficients of $\partial _t$ and $\partial _{\varphi}$ in (\ref{lambda_parti}) with those in (\ref{lambda_obs}) then yields
\begin{equation}\label{eq:alphaw}
\alpha =
\dfrac{
2 a p_{\varphi} - \omega _0 r_{\mathrm{O}}^3
}{
\sqrt{\Omega (r_{\mathrm{O}}, \vartheta _{\mathrm{O}} )} \, r_{\mathrm{O}}^3
} \, ,
\end{equation}
\begin{equation}\label{eq:betaw}
\beta =
\dfrac{p_{\varphi}
}{
\sqrt{\Omega (r_{\mathrm{O}}, \vartheta _{\mathrm{O}} )} \, r_{\mathrm{O}}
\, \mathrm{sin} \, \vartheta _{\mathrm{O}} \,
\mathrm{sin} \, \psi \, \mathrm{sin} \, \theta
} \, .
\end{equation}
Here we have used (\ref{eq:phdotw}) and (\ref{eq:tdotw}).
Similarly, we compare the coefficients of $\partial _r$ and $\partial _{\vartheta}$. Inserting the resulting expressions for $\dot{r}$ and $\dot{\vartheta}$ into (\ref{eq:rdotw}) and (\ref{eq:thdotw}), substituting for $\beta$ from (\ref{eq:betaw})  and solving for the celestial coordinates $\psi$ and $\theta$ gives:
\begin{equation}
    \mathrm{tan}^2\theta = \dfrac{\mathcal{K} - f_{\vartheta}(\vartheta _{\mathrm{O}})}{\left(\dfrac{2a}{r_{\mathrm{O}}^2}p _{\varphi} -\omega _0 r_{\mathrm{O}}\right)^2- \mathcal{K} - f_r ( r_{\mathrm{O}})},
\end{equation}

\begin{equation}
\mathrm{sin} ^2 \psi =
\dfrac{
p_{\varphi}^2
}{
\Big( \mathcal{K} - f_{\vartheta} ( \vartheta _{\mathrm{O}} ) \Big) \,
\mathrm{sin} ^2 \vartheta _{\mathrm{O}}
}    \, .
\end{equation}
Inserting $p_{\varphi} = p_{\varphi} (r_p)$ and $\mathcal{K} = \mathcal{K} (r_p)$ from (\ref{eq:paramw}) gives the part of the boundary curve of the shadow that is formed by light rays that spiral towards spherical light rays in the photon region; this curve is parametrized by $r_p$. Inserting $K= K ( p_{\varphi})$ from (\ref{eq:teoKp}) gives the part of the boundary curve of the shadow that corresponds to light rays that spiral towards the photon sphere at the neck; this curve is parametrized by $p_{\varphi}$.

\begin{figure}[h!]
    \centering
     \begin{tabular}{ll}
    (a) &  (b)\\
    \includegraphics[width=0.49\textwidth]{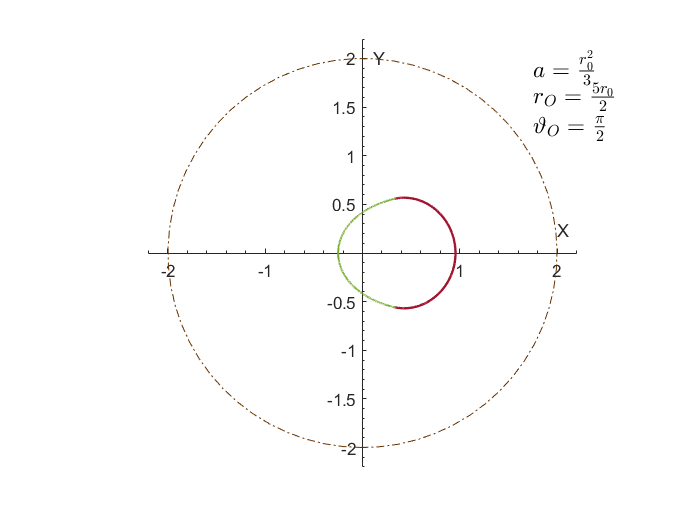} & \includegraphics[width=0.49\textwidth]{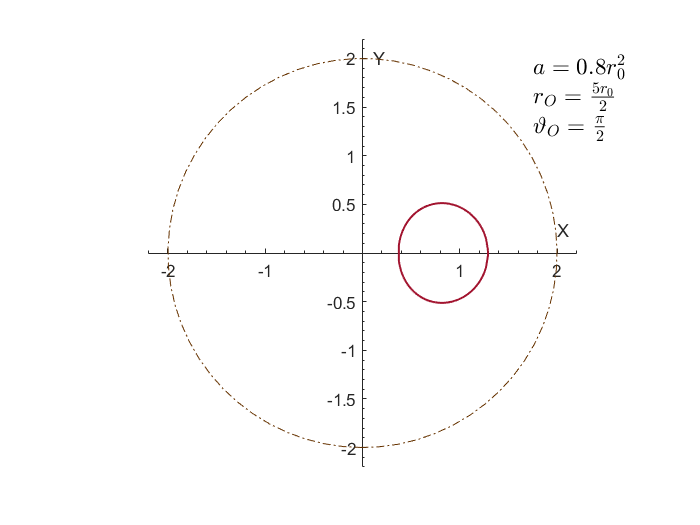}
  \end{tabular}
    \caption{Shadow of the Teo wormhole for an observer at $r_O=5r_0/2$ and $\vartheta_O=\pi/2$ in the case when (a) $a=r_0^2/3$, $f_r=0$, $f_\vartheta=0$  and (b) $a=0.8r_0^2$, $f_r=4\omega_0^2r_0^2(r_0/r)^{1/2}$, $f_\vartheta=0$. The thick purple curves show the boundary of the shadow that is given by light rays spiraling towards spherical light rays in the photon region, and the thin green curve corresponds to the shadow boundary formed by light rays that spiral towards the photon sphere at the neck, i.e., when $r=r_0$. The dash-dotted circle shows the position of the celestial equator.}
    \label{fig_02}
\end{figure}

The shadow as seen by an observer located at $r_O=5r_0/2$ and $\vartheta_O=\pi/2$ in vacuum and in a plasma, respectively, is shown in Fig.~\ref{fig_02}. The dimensionless Cartesian coordinates used in Ref.~\onlinecite{PerlickTsupko2017} were applied to depict the shadow curve. They read
\begin{gather}
    X(r)=-2\tan\left(\frac{\theta(r)}{2}\right)\sin(\psi(r)),\\
    Y(r)=-2\tan\left(\frac{\theta(r)}{2}\right)\cos(\psi(r)).
\end{gather}
The coordinates are defined in the plane that is tangent to the celestial sphere at the pole $\theta=0$ and the stereographic projection onto it is performed.

\section{Conclusions}
Axisymmetric stationary spacetimes represent a natural arena for objects which are of astrophysical interest. If such objects are surrounded by plasma, the light rays are affected by both curved geometry and properties of medium. We studied such situations starting from the elegant Hamiltonian formalism of the light rays propagating in general stationary axisymmetric spacetimes with   a non-magnetized (hence locally isotropic) and pressure-free plasma which is refractive and dispersive (Section II). If the plasma density shares the symmetry of the spacetime, the Killing vector fields associated with stationarity and axisymmetry give us two constants of motion in addition to the Hamiltonian. If the spacetime admits an equatorial plane, and if the plasma density is symmetric with respect to the equatorial plane, this gives us enough constants of motion for integrating the light rays in the equatorial plane, and also along the axis of symmetry. For general light rays, however, the equations of motion are not completely integrable. We investigated what conditions arise on the axisymmetric, stationary metric in the coordinates adapted to the symmetries (see Eq.~(\ref{metrika})) and   on the plasma density if we require the Hamilton-Jacobi equation for the rays to be separable. Similarly to the most relevant case of this type, the Kerr metric, another constant of motion, the ``generalized Carter constant'' must exist. In Section III we found these conditions (see Eqs.~(\ref{podm_BD}), (\ref{podm_omega}), (\ref{podm_ACP})). Assuming such constant to exist and employing the Hamilton equations of motion, we determined the photon region by calculating the position of the spherical light rays (Section IV) and the black hole shadow in a   stationary and axisymmetric spacetime with plasma (Section V). As the separability condition is local in nature, our results are applicable also to all other cases where we have two commuting Killing vector fields that span timelike two-surfaces. A noteworthy observation is that, on a spacetime with such symmetries, the separability condition cannot hold for light rays in any plasma density if it does not hold for light rays in vacuum.

After these considerations we analyzed several examples in Sections VI-IX.  First we showed how our general formulae work for the so-called  ``hairy Kerr metrics'' arising by assuming the mass in the Kerr metric to be a function of the radial (Boyer-Lindquist type) coordinate rather than a constant. Next we turned to the Hartle-Thorne approximate metric representing exterior regions of slowly rotating objects with a quadrupole moment   and verified with the help of our general results that in the chosen coordinates the full separability of variables in the Hamilton-Jacobi equation cannot be achieved unless in the limiting Schwarzschild case. An instructive example demonstrating the importance of the choice of a suitable coordinate system was presented in our discussion of the Melvin universe arising due to the strong ``uniform'' magnetic field; again we assumed also plasma to be present. The equations for rays are separable in the ``cylindrical-type'' coordinates, but not in the ``spherical-type'' coordinates.

The most detailed discussion was devoted to Teo's rotating traversable wormhole spacetime with plasma. The separability of  variables (the existence of the Carter constant) was shown to be possible if certain metric terms are functions only of the radial-type coordinate and if the plasma density is of a certain separated form. The photon region and the shadow of the wormhole with plasma were determined and the results were compared with those of Nedkova et al. \cite{NedkovaTinchevYazadjiev2013}. In Fig.~\ref{fig_01} the shapes of the photon regions around a specific Teo wormhole in vacuum and with some specific plasma distribution are constructed; in Fig.~\ref{fig_02} the shadows are compared.

Of course, there exist more cases of stationary and axisymmetric spacetimes with plasma in which the Carter-type constant will exist, and the separability of the Hamilton-Jacobi equation for rays will be feasible.
For example, the ``specific variant'' of the old Lense-Thirring metric (in their original asymptotic form) became recently of interest, see, e.g.,
Ref.~\onlinecite{Baines}, because they can be applied as a perfectly good approximation for the gravitational field generated by rotating sources with angular momentum.

\section*{Acknowledgments}
BB acknowledges the support of the Charles University Grant Agency under Contract No. 317421. BB and JB are also supported by the Czech Grant Agency under Contract No. 21/11268S. VP gratefully acknowledges support from Deutsche Forschungsgemeinschaft within the Research Training Group 1620 ``Models of Gravity.”

\section*{Appendix}
\setcounter{equation}{0}
\renewcommand{\theequation}{A\arabic{equation}}

In this appendix we   assume an axially symmetric and stationary metric, i.e., a metric given in coordinates $(t , \varphi , r , \vartheta )$ with the metric coefficients being independent of $t$ and $\varphi$, and we prove the following two statements: The Hamilton-Jacobi equation for light rays in a plasma can separate only if the plasma density is independent of $t$ and $\varphi$, and if separation holds in the chosen coordinates, then it holds in coordinates with $g_{tr}=g_{t \vartheta}=g_{\varphi r} = g_{\varphi \vartheta} = g_{r \vartheta}=0$, i.e., in coordinates such that the metric takes the form of Eq. (\ref{metrika}). To that end we assume that the two Killing vector fields, $\partial / \partial t$ and $\partial / \partial \varphi$, span two-dimensional hypersurfaces of signature $(+,-)$. This assumption implies that $g^{rr}$ and $g^{\vartheta \vartheta}$ are non-zero because a timelike vector cannot be tangential to a lightlike hypersurface.

We want to solve the Hamilton-Jacobi equation

\begin{equation}
    \mathcal{H} \Big( x^{\alpha}, \dfrac{\partial S}{\partial x ^{\beta}} \Big) = 0
\end{equation}
under the assumption that $S$ satisfies the separation form
\begin{equation}
S(x^{\alpha}) = S_t (t) + S _{\varphi} ( \varphi )
+ S_r (r) + S_{\vartheta} ( \vartheta ) \, .
\end{equation}
Then the Hamilton-Jacobi equation reads
\[
g^{tt} \Big( \dfrac{d S_t}{dt} \Big) ^2 + 2 g^{t \varphi}  \dfrac{d S_t }{dt} \, \dfrac{d S_{\varphi}}{d\varphi}   + g^{\varphi \varphi} \Big( \dfrac{d S_{\varphi}}{d\varphi} \Big) ^2 + 2 g^{tr} \dfrac{d S_t}{dt} \, \dfrac{d S_r}{dr} + 2 g^{t \vartheta} \, \dfrac{d S_t}{dt} \dfrac{dS_{\vartheta}}{d \vartheta} + 2 g^{\varphi r} \dfrac{d S_{\varphi}}{d\varphi} \, \dfrac{dS_r}{dr}
\]
\begin{equation}
+g^{\varphi \vartheta} \dfrac{d S_{\varphi}}{d\varphi} \, \dfrac{dS_{\vartheta}}{d \vartheta}+g^{rr} \Big( \dfrac{dS_r}{dr} \Big)^2 + 2 g^{r \vartheta} \dfrac{dS_r}{dr} \, \dfrac{dS_{\vartheta}}{d \vartheta} + g^{\vartheta \vartheta} \Big( \dfrac{d S_{\vartheta}}{d \vartheta} \Big)^2 + \omega _{\mathrm{pl}}^{\, 2}  = 0 \, .
\end{equation}
Differentiation with respect to $t$ yields
\begin{equation}
\Bigg( g^{tt} \dfrac{d S_t }{dt} + g^{t \varphi} \dfrac{d S_{\varphi}}{d \varphi} + g^{t r} \dfrac{d S_r}{d r} + g^{t \vartheta} \dfrac{d S_{\vartheta}}{d \vartheta} \Bigg)  \, \dfrac{d^2 S_t}{dt^2} + \omega _{\mathrm{pl}} \dfrac{\partial \omega _{\mathrm{pl}}}{\partial t} = 0 \, .
\end{equation}
As the plasma density is independent of the individual solution to the Hamilton-Jacobi equation, this can be true only if $\partial \omega _{\mathrm{pl}}/ \partial t$ is zero and $p_t = d S_t / dt$ is a constant. An analogous calculation shows that $\partial \omega _{\mathrm{pl}}/ \partial \varphi$ must be zero and $p_{\varphi} = d S_{\varphi} / d{\varphi}$ must be a constant, hence
\begin{equation}
S(x) = p_t \, t + p _{\varphi} \, \varphi
+ S_r (r) + S_{\vartheta} ( \vartheta )
\end{equation}
  with constants $p_t$ and $p_{\varphi}$.
Here we are free to multiply the left-hand side of the Hamilton-Jacobi equation with a function $F(r , \vartheta )$ that is non-zero but otherwise arbitrary. Written out in full, the Hamilton-Jacobi equation reads

\[
F \; \Bigg( g^{tt} p_t^2 + 2 g^{t \varphi} p_t p_{\varphi} + g^{\varphi \varphi} p_{\varphi} ^2 + 2 g^{tr} p_t \dfrac{d S_r}{dr} + 2 g^{t \vartheta} p_t \dfrac{dS_{\vartheta}}{d \vartheta} + 2 g^{\varphi r} p_{\varphi} \dfrac{dS_r}{dr}+g^{\varphi \vartheta} p_{\varphi} \dfrac{dS_{\vartheta}}{d \vartheta}
\]
\begin{equation}
+g^{rr} \Big( \dfrac{dS_r}{dr} \Big)^2 + 2 g^{r \vartheta} \dfrac{dS_r}{dr} \, \dfrac{dS_{\vartheta}}{d \vartheta} + g^{\vartheta \vartheta} \Big( \dfrac{d S_{\vartheta}}{d \vartheta} \Big)^2 + \omega _{\mathrm{pl}}^{\, 2} \Bigg) = 0 \, .
\end{equation}
Separability requires that the left-hand side is a funtion of $r$ only plus a function of $\vartheta$ only. As this has to hold for all $p_t$ and $p_{\varphi}$, and as the plasma density is independent of $p_t$, $p_\varphi$, $S_r$ and $S_{\vartheta}$, this   gives us the following set of equations.
\begin{equation}
F \, g^{tt} = \rho _r + \rho _{\vartheta}\, , \quad
F \, g^{t \varphi} = \lambda _r + \lambda _{\vartheta}\, , \quad
F \, g^{ \varphi \varphi} = \sigma _r + \sigma _{\vartheta}\, , \quad
\label{eq:sepcon3}
\end{equation}
\begin{equation}
F \, g^{tr} = \delta _r \, , \quad
F \, g^{t\vartheta} = \delta _{\vartheta} \, , \quad
F \, g^{\varphi r} = \varepsilon _r \, , \quad
F \, g^{\varphi \vartheta} = \varepsilon _{\vartheta} \, ,
\label{eq:sepcon1}
\end{equation}
\begin{equation}
F \, g^{rr} = \zeta _r \, , \quad
F \, g^{\vartheta \vartheta } = \zeta _{\vartheta} \, , \quad
g^{r \vartheta} = 0 \, .
\label{eq:sepcon2}
\end{equation}
\begin{equation}
F \, \omega _{\mathrm{pl}}^{\ , 2} = f _r + f _{\vartheta} \, .
\label{eq:sepcon4}
\end{equation}
Here and in the following, functions with an index $r$ depend on $r$ only and functions with an index $\vartheta$ depend on $\vartheta$ only. In particular we read from (\ref{eq:sepcon2}) that the separability can hold only if $g^{r \vartheta}=0$.

We now perform a coordinate transformation

\begin{equation}
t \mapsto t + \alpha _r + \alpha _{\vartheta} \, , \quad
\varphi  \mapsto \varphi + \beta _r + \beta _{\vartheta} \, , \quad
r \mapsto r \, , \quad
\vartheta \mapsto \vartheta \, ,
\end{equation}

\begin{equation}
dt \mapsto dt + \dfrac{d\alpha _r}{dr} \, dr + \dfrac{d \alpha _{\vartheta}}{d \vartheta} \, d \vartheta \, , \quad
d \varphi  \mapsto d \varphi + \dfrac{d \beta _r}{dr} \, dr  + \dfrac{d \beta _{\vartheta}}{d \vartheta} \, d \vartheta \, , \quad
d r \mapsto d r \, , \quad
d \vartheta \mapsto d \vartheta \, ,
\end{equation}

\begin{equation}
\dfrac{\partial}{\partial t} \mapsto \dfrac{\partial}{\partial t}  \, , \quad
\dfrac{\partial}{\partial \varphi} \mapsto \dfrac{\partial}{\partial \varphi}  \, , \quad
\dfrac{\partial}{\partial r} \mapsto \dfrac{\partial}{\partial r} - \dfrac{d \alpha _r}{dr} \, \dfrac{\partial}{\partial t} - \dfrac{d \beta _r}{d r} \, \dfrac{\partial}{\partial \varphi} \, , \quad
\dfrac{\partial}{\partial \vartheta} \mapsto \dfrac{\partial}{\partial \vartheta} - \dfrac{d \alpha _{\vartheta}}{d \vartheta} \, \dfrac{\partial}{\partial t} - \dfrac{d \beta _{\vartheta}}{d \vartheta} \, \dfrac{\partial}{\partial \varphi} \, ,
\end{equation}
hence

\begin{equation}
g^{tr} \mapsto g^{tr} + \dfrac{d\alpha _r}{dr} \, g^{rr} + \dfrac{d \alpha _{\vartheta}}{d \vartheta} \, g^{\vartheta r} = \dfrac{1}{F} \Bigg( \delta _r + \dfrac{d \alpha _r}{dr} \, \zeta _r \Bigg) \, ,
\end{equation}
\begin{equation}
g^{t \vartheta} \mapsto g^{t \vartheta} + \dfrac{d\alpha _r}{dr} \, g^{r \vartheta} + \dfrac{d \alpha _{\vartheta}}{d \vartheta} \, g^{\vartheta  \vartheta} = \dfrac{1}{F} \Bigg( \delta _{\vartheta} + \dfrac{d \alpha _{\vartheta}}{d \vartheta} \, \zeta _{\vartheta} \Bigg) \, , \end{equation}
\begin{equation}
g^{\varphi r}  \mapsto g^{\varphi r} + \dfrac{d \beta _r}{dr} \, g^{rr}  + \dfrac{d \beta _{\vartheta}}{d \vartheta} \, g^{\vartheta r} = \dfrac{1}{F} \Bigg( \varepsilon _r + \dfrac{d \beta _r}{r} \, \zeta _r \Bigg) \, ,
\end{equation}
\begin{equation}
g^{\varphi \vartheta}  \mapsto g^{\varphi \vartheta} + \dfrac{d \beta _r}{dr} \, g^{r \vartheta}  + \dfrac{d \beta _{\vartheta}}{d \vartheta} \, g^{\vartheta \vartheta} = \dfrac{1}{F} \Bigg( \varepsilon _{\vartheta} + \dfrac{d \beta  _{\vartheta}}{d \vartheta} \, \zeta _{\vartheta} \Bigg) \, .
\end{equation}
As $\zeta _r= F g^{rr}$ and $\zeta _{\vartheta} = F g^{\vartheta \vartheta}$ are non-zero, we can choose functions $\alpha _r$, $\alpha _{\vartheta}$, $\beta _r$ and $\beta _{\vartheta}$ such that the right-hand sides of these four equations are zero, i.e., such that in the new coordinates the metric components $g^{tr}$, $g^{t \vartheta}$, $g^{\varphi r}$ and $g^{\varphi \vartheta}$ vanish. The conditions (\ref{eq:sepcon1}) are then still satisfied, now with the right-hand sides equal to zero. Eqs. (\ref{eq:sepcon2}) and (\ref{eq:sepcon4}) are unchanged, whereas Eqs. (\ref{eq:sepcon3}) are still true, but now with new functions $\rho _r$, $\rho _{\vartheta}$, $\lambda _{r}$, $\lambda _{\vartheta}$, $\sigma _r$, and $\sigma _{\vartheta}$, so separability still holds in the new coordinates if it did so in the original coordinates. By inverting the matrix $(g^{\mu \nu})$ we see that in the new coordinates the metric takes indeed the form of Eq. (\ref{metrika}).

\bibliographystyle{spphys}

\end{document}